\pdfoutput=1
\documentclass[12pt,a4paper]{article}

\usepackage{ifthen} % for conditional statements
\newboolean{pdflatex}
\setboolean{pdflatex}{true} % False for eps figures 

\newboolean{articletitles}
\setboolean{articletitles}{true} % False removes titles in references

\newboolean{uprightparticles}
\setboolean{uprightparticles}{false} %True for upright particle symbols

\newboolean{inbibliography}
\setboolean{inbibliography}{false} %True once you enter the bibliography

\usepackage{microtype}
\usepackage{lineno}  % for line numbering during review
\usepackage{xspace} % To avoid problems with missing or double spaces after
\usepackage{caption} %these three command get the figure and table captions automatically small

\usepackage{graphicx}  % to include figures (can also use other packages)
\usepackage{color}
\usepackage{colortbl}
\graphicspath{{./figs/}} % Make Latex search fig subdir for figures

\usepackage{amsmath} % Adds a large collection of math symbols
\usepackage{amssymb}
\usepackage{amsfonts}
\usepackage{upgreek} % Adds in support for greek letters in roman typeset

\newcommand*\patchAmsMathEnvironmentForLineno[1]{%
\expandafter\let\csname old#1\expandafter\endcsname\csname #1\endcsname
\expandafter\let\csname oldend#1\expandafter\endcsname\csname
end#1\endcsname
 \renewenvironment{#1}%
   {\linenomath\csname old#1\endcsname}%
   {\csname oldend#1\endcsname\endlinenomath}%
}
\newcommand*\patchBothAmsMathEnvironmentsForLineno[1]{%
  \patchAmsMathEnvironmentForLineno{#1}%
  \patchAmsMathEnvironmentForLineno{#1*}%
}
\AtBeginDocument{%
\patchBothAmsMathEnvironmentsForLineno{equation}%
\patchBothAmsMathEnvironmentsForLineno{align}%
\patchBothAmsMathEnvironmentsForLineno{flalign}%
\patchBothAmsMathEnvironmentsForLineno{alignat}%
\patchBothAmsMathEnvironmentsForLineno{gather}%
\patchBothAmsMathEnvironmentsForLineno{multline}%
\patchBothAmsMathEnvironmentsForLineno{eqnarray}%
}

\usepackage{hyperref}    % Hyperlinks in references
\usepackage[all]{hypcap} % Internal hyperlinks to floats.

\def\lhcb {\mbox{LHCb}\xspace}

\def\MagUp {\mbox{\em Mag\kern -0.05em Up}\xspace}

\ifthenelse{\boolean{uprightparticles}}%
{

 \def\Pmu         {\ensuremath{\upmu}\xspace}                 
 \def\Pnu         {\ensuremath{\upnu}\xspace}                 
                  
 \def\Ppi         {\ensuremath{\uppi}\xspace}

 \def\Ptau        {\ensuremath{\uptau}\xspace}

 \def\Ppsi        {\ensuremath{\uppsi}\xspace}

 \def\PDelta      {\ensuremath{\Delta}\xspace}                 
 \def\PXi      {\ensuremath{\Xi}\xspace}                 
 \def\PLambda      {\ensuremath{\Lambda}\xspace}                 
 \def\PSigma      {\ensuremath{\Sigma}\xspace}                 
 \def\POmega      {\ensuremath{\Omega}\xspace}                 
 \def\PUpsilon      {\ensuremath{\Upsilon}\xspace}                 
 
 \def\PB      {\ensuremath{\mathrm{B}}\xspace}                 
                  
 \def\PD      {\ensuremath{\mathrm{D}}\xspace}

 \def\PJ      {\ensuremath{\mathrm{J}}\xspace}                 
 \def\PK      {\ensuremath{\mathrm{K}}\xspace}

 \def\PX      {\ensuremath{\mathrm{X}}\xspace}

 \def\Pb      {\ensuremath{\mathrm{b}}\xspace}                 
 \def\Pc      {\ensuremath{\mathrm{c}}\xspace}

 \def\Pi      {\ensuremath{\mathrm{i}}\xspace}

}
{

 \def\Pmu         {\ensuremath{\mu}\xspace}                 
 \def\Pnu         {\ensuremath{\nu}\xspace}                 
                  
 \def\Ppi         {\ensuremath{\pi}\xspace}

 \def\Ptau        {\ensuremath{\tau}\xspace}

 \def\Ppsi        {\ensuremath{\psi}\xspace}                 
                  
 \mathchardef\PDelta="7101
 \mathchardef\PXi="7104
 \mathchardef\PLambda="7103
 \mathchardef\PSigma="7106
 \mathchardef\POmega="710A
 \mathchardef\PUpsilon="7107
                  
 \def\PB      {\ensuremath{B}\xspace}                 
                  
 \def\PD      {\ensuremath{D}\xspace}

 \def\PJ      {\ensuremath{J}\xspace}                 
 \def\PK      {\ensuremath{K}\xspace}

 \def\PX      {\ensuremath{X}\xspace}

 \def\Pb      {\ensuremath{b}\xspace}                 
 \def\Pc      {\ensuremath{c}\xspace}

 \def\Pi      {\ensuremath{i}\xspace}

}

\makeatletter
\ifcase \@ptsize \relax% 10pt
  \newcommand{\miniscule}{\@setfontsize\miniscule{4}{5}}% \tiny: 5/6
\or% 11pt
  \newcommand{\miniscule}{\@setfontsize\miniscule{5}{6}}% \tiny: 6/7
\or% 12pt
  \newcommand{\miniscule}{\@setfontsize\miniscule{5}{6}}% \tiny: 6/7
\fi
\makeatother

\DeclareRobustCommand{\optbar}[1]{\shortstack{{\miniscule (\rule[.5ex]{1.25em}{.18mm})}
  \\ [-.7ex] $#1$}}

\def\mun        {{\ensuremath{\Pmu^-}}\xspace} % muon negative (\mum is taken)

\def\neub       {{\ensuremath{\overline{\Pnu}}}\xspace}

\def\neumb      {{\ensuremath{\neub_\mu}}\xspace}
\def\cquark    {{\ensuremath{\Pc}}\xspace}

\def\bquark    {{\ensuremath{\Pb}}\xspace}

\def\pion   {{\ensuremath{\Ppi}}\xspace}

\def\pip    {{\ensuremath{\pion^+}}\xspace}
\def\pim    {{\ensuremath{\pion^-}}\xspace}

\def\kaon    {{\ensuremath{\PK}}\xspace}
  \def\Kbar    {{\kern 0.2em\overline{\kern -0.2em \PK}{}}\xspace}

\def\KorKbar    {\kern 0.18em\optbar{\kern -0.18em K}{}\xspace}

\def\Kp      {{\ensuremath{\kaon^+}}\xspace}
\def\Km      {{\ensuremath{\kaon^-}}\xspace}

  \def\Dbar    {{\kern 0.2em\overline{\kern -0.2em \PD}{}}\xspace}
\def\D       {{\ensuremath{\PD}}\xspace}

\def\DorDbar    {\kern 0.18em\optbar{\kern -0.18em D}{}\xspace}
\def\Dz      {{\ensuremath{\D^0}}\xspace}
\def\Dzb     {{\ensuremath{\Dbar{}^0}}\xspace}

\def\Dm      {{\ensuremath{\D^-}}\xspace}

\def\Dstarp  {{\ensuremath{\D^{*+}}}\xspace}

\def\B       {{\ensuremath{\PB}}\xspace}
\def\Bbar    {{\ensuremath{\kern 0.18em\overline{\kern -0.18em \PB}{}}}\xspace}

\def\BorBbar    {\kern 0.18em\optbar{\kern -0.18em B}{}\xspace}

\def\Bzb     {{\ensuremath{\Bbar{}^0}}\xspace}

\def\Bub     {{\ensuremath{\B^-}}\xspace}

\def\Bm      {{\ensuremath{\Bub}}\xspace}

\def\Bd      {{\ensuremath{\B^0}}\xspace}

\def\Bdb     {{\ensuremath{\Bbar{}^0}}\xspace}

\def\jpsi     {{\ensuremath{{\PJ\mskip -3mu/\mskip -2mu\Ppsi\mskip 2mu}}}\xspace}
\def\psitwos  {{\ensuremath{\Ppsi{(2S)}}}\xspace}

  \def\Y#1S{\ensuremath{\PUpsilon{(#1S)}}\xspace}% no space before {...}!

\def\Lbar        {{\ensuremath{\kern 0.1em\overline{\kern -0.1em\PLambda}}}\xspace}
\def\LorLbar    {\kern 0.18em\optbar{\kern -0.18em \PLambda}{}\xspace}

\newcommand{\decay}[2]{\ensuremath{#1\!\to #2}\xspace}         % {\Pa}{\Pb \Pc}

\def\to                 {\ensuremath{\rightarrow}\xspace}

\def\order   {{\ensuremath{\mathcal{O}}}\xspace}

\def\CP                {{\ensuremath{C\!P}}\xspace}

\newcommand{\Adir}{{\ensuremath{{\cal A}^{\rm dir}}}\xspace}
\newcommand{\Amix}{{\ensuremath{{\cal A}^{\rm mix}}}\xspace}

\newcommand{\mistag}{\ensuremath{\omega}\xspace}

\def\AT#1     {\ensuremath{A_{\mathrm{T}}^{#1}}\xspace}           % 2

\def\C#1      {\ensuremath{\mathcal{C}_{#1}}\xspace}                       % 9
\def\Cp#1     {\ensuremath{\mathcal{C}_{#1}^{'}}\xspace}                    % 7
\def\Ceff#1   {\ensuremath{\mathcal{C}_{#1}^{\mathrm{(eff)}}}\xspace}        % 9  
\def\Cpeff#1  {\ensuremath{\mathcal{C}_{#1}^{'\mathrm{(eff)}}}\xspace}       % 7
\def\Ope#1    {\ensuremath{\mathcal{O}_{#1}}\xspace}                       % 2
\def\Opep#1   {\ensuremath{\mathcal{O}_{#1}^{'}}\xspace}                    % 7

\newcommand{\ket}[1]{\ensuremath{|#1\rangle}}              % {b}
\newcommand{\tev}{\ifthenelse{\boolean{inbibliography}}{\ensuremath{~T\kern -0.05em eV}\xspace}{\ensuremath{\mathrm{\,Te\kern -0.1em V}}}\xspace}
\newcommand{\gev}{\ensuremath{\mathrm{\,Ge\kern -0.1em V}}\xspace}
\newcommand{\mev}{\ensuremath{\mathrm{\,Me\kern -0.1em V}}\xspace}
\newcommand{\kev}{\ensuremath{\mathrm{\,ke\kern -0.1em V}}\xspace}
\newcommand{\ev}{\ensuremath{\mathrm{\,e\kern -0.1em V}}\xspace}
\newcommand{\gevc}{\ensuremath{{\mathrm{\,Ge\kern -0.1em V\!/}c}}\xspace}
\newcommand{\mevc}{\ensuremath{{\mathrm{\,Me\kern -0.1em V\!/}c}}\xspace}
\newcommand{\gevcc}{\ensuremath{{\mathrm{\,Ge\kern -0.1em V\!/}c^2}}\xspace}
\newcommand{\gevgevcccc}{\ensuremath{{\mathrm{\,Ge\kern -0.1em V^2\!/}c^4}}\xspace}
\newcommand{\mevcc}{\ensuremath{{\mathrm{\,Me\kern -0.1em V\!/}c^2}}\xspace}

\def\mum  {\ensuremath{{\,\upmu\rm m}}\xspace}

\def\invfb   {\ensuremath{\mbox{\,fb}^{-1}}\xspace}

\def\ps   {\ensuremath{{\rm \,ps}}\xspace}
\def\fs   {\ensuremath{\rm \,fs}\xspace}

\def\order{{\ensuremath{\cal O}}\xspace}
\newcommand{\chisq}{\ensuremath{\chi^2}\xspace}

\def\gsim{{~\raise.15em\hbox{$>$}\kern-.85em
          \lower.35em\hbox{$\sim$}~}\xspace}
\def\lsim{{~\raise.15em\hbox{$<$}\kern-.85em
          \lower.35em\hbox{$\sim$}~}\xspace}

\def\ptot       {\mbox{$p$}\xspace}
\def\pt         {\mbox{$p_{\rm T}$}\xspace}

\def\evtgen     {\mbox{\textsc{EvtGen}}\xspace}

\def\geant      {\mbox{\textsc{Geant4}}\xspace}

\def\photos     {\mbox{\textsc{Photos}}\xspace}

\def\pythia     {\mbox{\textsc{Pythia}}\xspace}

\def\tell1  {TELL1\xspace}
\def\ukl1   {UKL1\xspace}

\newcommand{\eg}{\mbox{\itshape e.g.}\xspace}
\newcommand{\ie}{\mbox{\itshape i.e.}\xspace}

\usepackage{cite} % Allows for ranges in citations
\usepackage{mciteplus}

\def\AGKKval {-0.134}
\def\AGKKstat {0.077}
\def\AGKKsystP {0.026}
\def\AGKKsystM {0.034}
\def\AGKKscale {0.10}

\def\AGpipival {-0.092}
\def\AGpipistat {0.145}
\def\AGpipisystP {0.025}
\def\AGpipisystM {0.033}
\def\AGpipiscale {0.09}

\def\AGKpival {0.009}
\def\AGKpistat {0.032}

\newcommand{\AG}{\ensuremath{A_{\Gamma}}\xspace}
\newcommand{\AGKK}{\ensuremath{A_{\Gamma}(\Km\Kp)}\xspace}
\newcommand{\AGpipi}{\ensuremath{A_{\Gamma}(\pim\pip)}\xspace}
\newcommand{\AGKpi}{\ensuremath{A_{\Gamma}(\Km\pip)}\xspace}
\newcommand{\ACP}{\ensuremath{A_{\CP}}\xspace}

\def\Adir   {\ensuremath{A_{\CP}^{\rm dir}}\xspace}
\def\Amix   {\ensuremath{A_{\CP}^{\rm mix}}\xspace}

\def\dkk        {\decay{\Dz}{\Km\Kp}}
\def\dpipi      {\decay{\Dz}{\pim\pip}}
\def\dkpi       {\decay{\Dz}{\Km\pip}}
\def\dkpiDCS    {\decay{\Dz}{\Kp\pim}}

\newcommand{\gevcNS}{\ensuremath{{\mathrm{Ge\kern -0.1em V\!/}c}}\xspace}

\DeclareGraphicsExtensions{.pdf,.PDF,png,.PNG}

\begin{document}

%%%%%%%%%%%%%%%%%%%%%%%%%
%%%%% Title     %%%%%%%%%
%%%%%%%%%%%%%%%%%%%%%%%%%
\renewcommand{\thefootnote}{\fnsymbol{footnote}}
\setcounter{footnote}{1}

% %%%%%%% CHOOSE TITLE PAGE--------
% $Id: title-LHCb-PAPER.tex 61931 2014-10-14 09:51:37Z roldeman $
% ===============================================================================
% Purpose: LHCb-PAPER journal paper title page template
% Author: 
% Created on: 2010-09-25
% ===============================================================================

%%%%%%%%%%%%%%%%%%%%%%%%%
%%%%%  TITLE PAGE  %%%%%%
%%%%%%%%%%%%%%%%%%%%%%%%%
\begin{titlepage}
\pagenumbering{roman}

% Header ---------------------------------------------------
\vspace*{-1.5cm}
\centerline{\large EUROPEAN ORGANIZATION FOR NUCLEAR RESEARCH (CERN)}
\vspace*{1.5cm}
\hspace*{-0.5cm}
\begin{tabular*}{\linewidth}{lc@{\extracolsep{\fill}}r}
\ifthenelse{\boolean{pdflatex}}% Logo format choice
{\vspace*{-2.7cm}\mbox{\!\!\!\includegraphics[width=.14\textwidth]{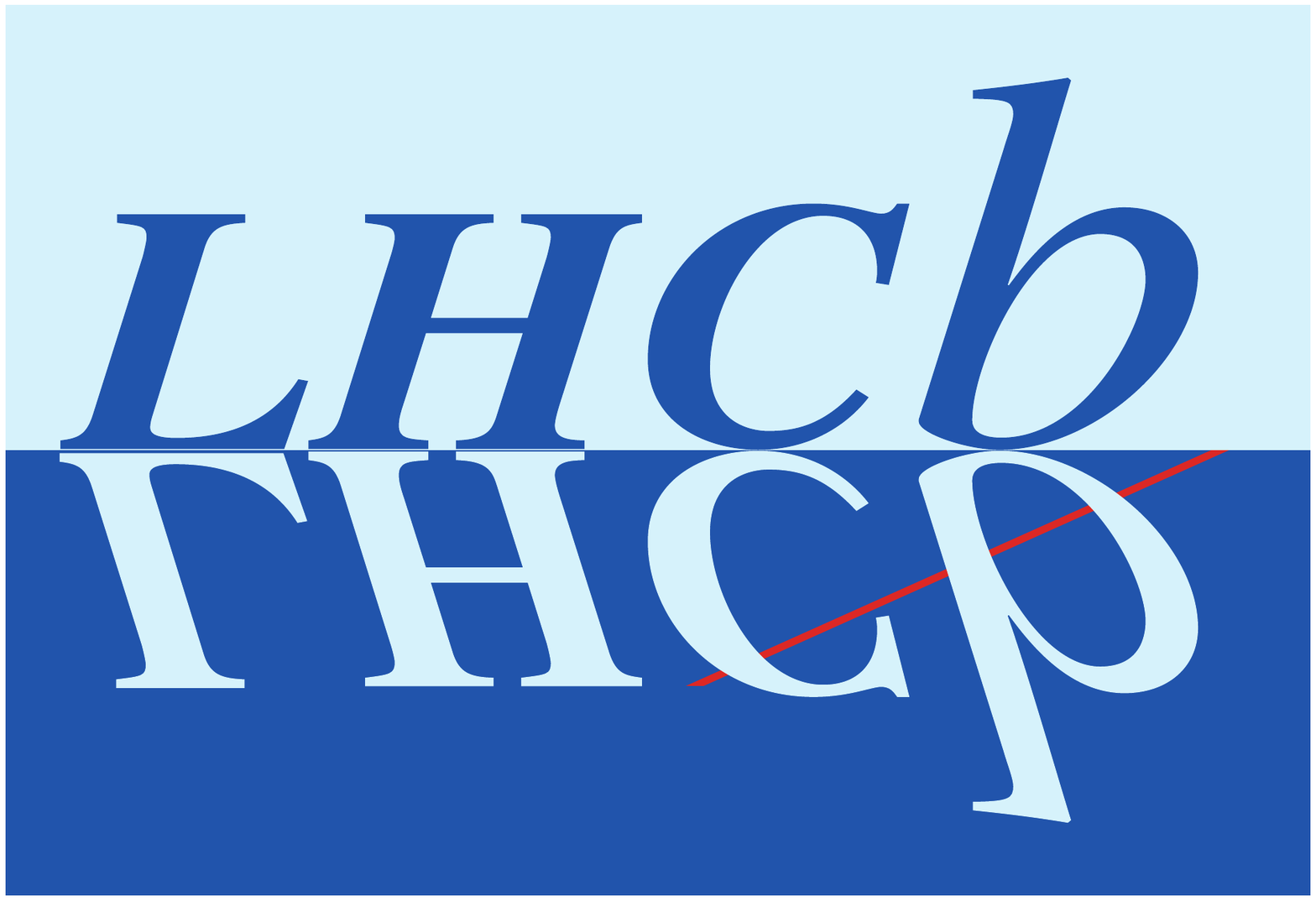}} & &}%
{\vspace*{-1.2cm}\mbox{\!\!\!\includegraphics[width=.12\textwidth]{lhcb-logo.eps}} & &}%
\\
 & & CERN-PH-EP-2015-008 \\  % ID 
 & & LHCb-PAPER-2014-069 \\  % ID 
 & & April 13, 2015 \\
 & & \\
\end{tabular*}

\vspace*{1.0cm}

% Title --------------------------------------------------
{\bf\boldmath\huge
\begin{center}
  Measurement of indirect \CP asymmetries in \dkk and \dpipi decays using semileptonic \PB decays
\end{center}
}

\vspace*{1.0cm}

% Authors -------------------------------------------------
\begin{center}
The LHCb collaboration\footnote{Authors are listed at the end of this paper.}
\end{center}

\vspace{\fill}

% Abstract -----------------------------------------------
\begin{abstract}
  \noindent
Time-dependent \CP asymmetries in the decay rates of the singly
Cabibbo-suppressed decays \dkk and \dpipi are measured in $pp$ collision data
corresponding to an integrated luminosity of 3.0\invfb collected by the LHCb
experiment. The \Dz mesons are produced in semileptonic \bquark-hadron decays,
where the charge of the accompanying muon is used to determine the initial state
as \Dz or \Dzb. The asymmetries in effective lifetimes between \Dz and \Dzb
decays, which are sensitive to indirect \CP violation, are determined to be
\begin{align*}
  \AGKK  = (\AGKKval \pm \AGKKstat \; {}^{+\AGKKsystP}_{-\AGKKsystM})\% \ , \\
  \AGpipi = (\AGpipival\pm \AGpipistat \; {}^{+\AGpipisystP}_{-\AGpipisystM})\% \ ,
\end{align*}
where the first uncertainties are statistical and the second systematic. This
result is in agreement with previous measurements and with the hypothesis of no
indirect \CP violation in \Dz decays.
\end{abstract}

\vspace*{1.0cm}

\begin{center}
  Submitted to JHEP
\end{center}

\vspace{\fill}

{\footnotesize 
\centerline{\copyright~CERN on behalf of the \lhcb collaboration, licence \href{http://creativecommons.org/licenses/by/4.0/}{CC-BY-4.0}.}}
\vspace*{2mm}

\end{titlepage}

%%%%%%%%%%%%%%%%%%%%%%%%%%%%%%%%
%%%%%  EOD OF TITLE PAGE  %%%%%%
%%%%%%%%%%%%%%%%%%%%%%%%%%%%%%%%

%  empty page follows the title page ----
\newpage
\setcounter{page}{2}
\mbox{~}

\cleardoublepage

% %%%%%%%%%%%%% ---------

\renewcommand{\thefootnote}{\arabic{footnote}}
\setcounter{footnote}{0}

%%%%%%%%%%%%%%%%%%%%%%%%%
%%%%% Main text %%%%%%%%%
%%%%%%%%%%%%%%%%%%%%%%%%%

\pagestyle{plain} % restore page numbers for the main text
\setcounter{page}{1}
\pagenumbering{arabic}

\section{Introduction}
\label{sec:Introduction}

In neutral meson systems, mixing may occur between the particle and
anti-particle states. This mixing is very small in the charm-meson (\Dz)
system. Experimentally, a small, non-zero \Dz--\Dzb mixing is now firmly
established by several experiments~\cite{Lees:2012qh, delAmoSanchez:2010xz,
  LHCb-PAPER-2013-053, Aaltonen:2013pja, Ko:2014qvu, Peng:2014oda}, where the
average of these measurements excludes zero mixing at more than 11 standard
deviations~\cite{HFAG}.  This opens up the possibility to search for a breaking
of the charge-parity (\CP) symmetry occurring in the \Dz--\Dzb mixing alone or
in the interference between the mixing and decay amplitudes. This is called
indirect \CP violation and the corresponding asymmetry is predicted to be
$\order(10^{-4})$~\cite{Bianco:2003vb, Bobrowski:2010xg}, but can be enhanced in
theories beyond the Standard Model~\cite{Grossman:2006jg}. Indirect \CP
violation can be measured in decays to \CP eigenstates such as the singly
Cabibbo-suppressed decays \dkk and \dpipi (the inclusion of charge-conjugate
processes is implied hereafter) from the asymmetry between the effective \Dz and
\Dzb lifetimes, $\AG$. The effective lifetime is the lifetime obtained from a
single exponential fit to the decay-time distribution. Several measurements of
\AG exist~\cite{Lees:2012qh, LHCb-PAPER-2013-054, Aaltonen:2014efa}. The most
precise determination was made by \lhcb with data corresponding to 1.0\invfb of
integrated luminosity, resulting in $\AGKK = (-0.035\pm0.062\pm0.012)\%$, and
$\AGpipi = (0.033\pm0.106\pm0.014)\%$~\cite{LHCb-PAPER-2013-054}.  When indirect
\CP violation is assumed to be the same in the two modes, the world average
becomes $\AG = (-0.014\pm0.052)\%$~\cite{HFAG}. In all previous measurements of
\AG, the initial flavour of the neutral charm meson (\ie, whether it was a \Dz
or \Dzb state) was determined (tagged) by the charge of the pion in a
$\Dstarp\to\Dz\pip$ decay. In this paper, the time-dependent \CP asymmetry is
measured in \Dz decays originating from semileptonic \bquark-hadron decays,
where the charge of the accompanying muon is used to tag the flavour of the \Dz
meson. These samples are dominated by $\Bm\to\Dz\mun\neumb\PX$ and
$\Bzb\to\Dz\mun\neumb\PX$ decays, where \PX denotes other particle(s) possibly
produced in the decay. The same data samples as for the measurement of
time-integrated \CP asymmetries~\cite{LHCb-PAPER-2014-013} are used.

\section{Formalism and method}
\label{sec:method}

The time-dependent \CP asymmetry for a neutral \D meson decaying to a \CP
eigenstate, $f$, is defined as
\begin{align}
  \ACP(t) \equiv \frac{\Gamma(\Dz\to f;t) - \Gamma(\Dzb\to f;t)}
      {\Gamma(\Dz\to f;t) + \Gamma(\Dzb\to f;t)} \ ,
\end{align}
where $\Gamma(\Dz\to f;t)$ and $\Gamma(\Dzb\to f;t)$ are the time-dependent
partial widths of initial \Dz and \Dzb mesons to final state $f$.  The \CP
asymmetry can be approximated as~\cite{Aaltonen:2011se}
\begin{align}
  \ACP(t) \approx \Adir - \AG \frac{t}{\tau} \ ,
  \label{eq:ACPt}
\end{align}
where \Adir is the direct \CP asymmetry and $\tau$ is the \Dz lifetime. The
linear decay-time dependence is determined by \AG, which is formally defined as
\begin{align}
\AG \equiv \frac{\hat{\Gamma}_{\Dz} -\hat{\Gamma}_{\Dzb}}
    {\hat{\Gamma}_{\Dz} + \hat{\Gamma}_{\Dzb}} \ ,
\end{align}
where $\hat{\Gamma}$ is the effective partial decay rate of an initial
\Dz or \Dzb state to the \CP eigenstate. Furthermore, \AG can be approximated in
terms of the \Dz--\Dzb mixing parameters, $x$ and $y$,
as~\cite{Gersabeck:2011xj}
\begin{align}
  \AG \approx (\Amix/2 - \Adir)\,y\,\cos\phi - x \, \sin\phi \ ,
\end{align}
where $\Amix=|q/p|^2-1$ describes \CP violation in \Dz--\Dzb mixing, with $q$
and $p$ the coefficients of the transformation from the flavour basis to the
mass basis, $\ket{D_{1,2}}=p\ket{\Dz}\pm q\ket{\Dzb}$. The weak phase $\phi$
describes \CP violation in the interference between mixing and decay, and is
specific to the decay mode. Finally, \AG receives a contribution from direct \CP
violation as well~\cite{Kagan:2009gb}.

The raw asymmetry is affected by the different detection efficiencies
for positive and negative muons, and the different production rates of \Dz and
\Dzb mesons. These effects introduce a shift to the constant term in
Eq.~(\ref{eq:ACPt}), but have a negligible effect on the measurement of \AG (see
Sect.~\ref{sec:systematics}). The decay \dkpi, also flavour-tagged by the muon
from a semileptonic \bquark-hadron decay, is used as a control channel. Since
this is a Cabibbo-favoured decay mode, direct \CP violation is expected to be
negligible. More importantly, any indirect \CP violation is heavily suppressed
as the contribution from doubly Cabibbo-suppressed \dkpiDCS decays is small.

\section{Detector and simulation}
\label{sec:Detector}

The \lhcb detector~\cite{Alves:2008zz,LHCb-DP-2014-002} is a single-arm forward
spectrometer covering the \mbox{pseudorapidity} range $2<\eta <5$, designed for
the study of particles containing \bquark or \cquark quarks. The detector
includes a high-precision tracking system consisting of a silicon-strip vertex
detector surrounding the $pp$ interaction region, a large-area silicon-strip
detector located upstream of a dipole magnet with a bending power of about
$4{\rm\,Tm}$, and three stations of silicon-strip detectors and straw drift
tubes placed downstream of the magnet. The polarity of the magnetic field is
regularly reversed during data taking. The tracking system provides a
measurement of momentum, \ptot, with a relative uncertainty that varies from
0.5\% at low momentum to 1.0\% at 200\gevc. The minimum distance of a track to a
primary vertex, the impact parameter, is measured with a resolution of
$(15+29/\pt)\mum$, where \pt is the component of the momentum transverse to the
beam, in \gevcNS.  Different types of charged hadrons are distinguished using
information from two ring-imaging Cherenkov detectors. Photon, electron and
hadron candidates are identified by a calorimeter system consisting of
scintillating-pad and preshower detectors, an electromagnetic calorimeter and a
hadronic calorimeter. Muons are identified by a system composed of alternating
layers of iron and multiwire proportional chambers, situated behind the hadronic
calorimeter. The trigger~\cite{LHCb-DP-2012-004} consists of a hardware stage,
based on information from the calorimeter and muon systems, followed by a
software stage, which applies a full event reconstruction.

In the simulation, $pp$ collisions are generated using
\pythia~\cite{Sjostrand:2006za,*Sjostrand:2007gs} with a specific \lhcb
configuration~\cite{LHCb-PROC-2010-056}. Decays of hadronic particles are
described by \evtgen~\cite{Lange:2001uf}, in which final-state radiation is
generated using \photos~\cite{Golonka:2005pn}. The interaction of the generated
particles with the detector, and its response, are implemented using the \geant
toolkit~\cite{Allison:2006ve, *Agostinelli:2002hh} as described in
Ref.~\cite{LHCb-PROC-2011-006}.

\section{Data set and selection}
\label{sec:selection}

This analysis uses a data set corresponding to an integrated luminosity of
$3.0\invfb$.  The data were taken at two different $pp$ centre-of-mass energies:
7\tev in 2011 ($1.0\invfb$) and 8\tev in 2012 ($2.0\invfb$). The data sets
recorded with each dipole magnet polarity are roughly equal in size.

At the hardware trigger stage, the events are triggered by the presence of the
muon candidate in the muon system. This requires the muon \pt to be greater than
$1.64\gevc$ ($1.76\gevc$) for the 2011 (2012) data. At the software trigger
stage, one of the final-state particles is required to have enough momentum and
be significantly displaced from any primary $pp$ vertex. In addition, the
candidates must be selected by a single-muon trigger or by a topological trigger
that requires the muon and one or two of the \Dz daughters to be consistent with
the topology of \bquark-hadron decays~\cite{LHCb-DP-2012-004}.

To further suppress background, the \Dz daughters are required to have
$\pt>300\mevc$. All final-state particles are required to have a large impact
parameter and be well identified by the particle identification systems. The
impact parameter requirement on the muon reduces the contribution from \Dz
mesons produced directly in the $pp$ interaction to below $2\%$. The scalar \pt
sum of the \Dz daughters should be larger than $1.4\gevc$, and the \pt of the
\Dz candidate should be larger than $0.5\gevc$. The two tracks from the \Dz
candidate and the $\Dz\mu$ combination are required to form good vertices and
the latter vertex should be closer to the primary vertex than the \Dz
vertex. The \Dz decay time is determined from the distance between these two
vertices, and the reconstructed \Dz momentum.  The invariant mass of the
$\Dz\mu$ combination is required to be between $2.5$ and $5.0\gevcc$, where the
upper bound suppresses hadronic \bquark-hadron decays into three-body final
states. Backgrounds from inclusive \bquark-hadron decays into charmonium are
suppressed by vetoing candidates where the invariant mass of the muon and the
oppositely charged \Dz daughter, misidentified as a muon, is consistent with the
\jpsi or \psitwos mass. Additionally, the invariant mass of the muon and
same-charge \Dz daughter, under the muon mass hypothesis, is required to be
larger than $240\mevcc$ to remove events where a single charged particle is
reconstructed as two separate tracks. For most selection requirements, the
efficiency is roughly independent of the \Dz decay time, giving efficiency
variations of $\order(1\%)$. The largest dependence on the decay time comes from
the topological trigger, which introduces an efficiency profile that decreases
with \Dz decay time, resulting in about 20\% relative efficiency loss at large
decay times.

\section[Determination of AGamma]{Determination of {\boldmath \AG}}
\label{sec:results}

\begin{figure}
  \begin{center}
    \includegraphics[width=0.49\textwidth]{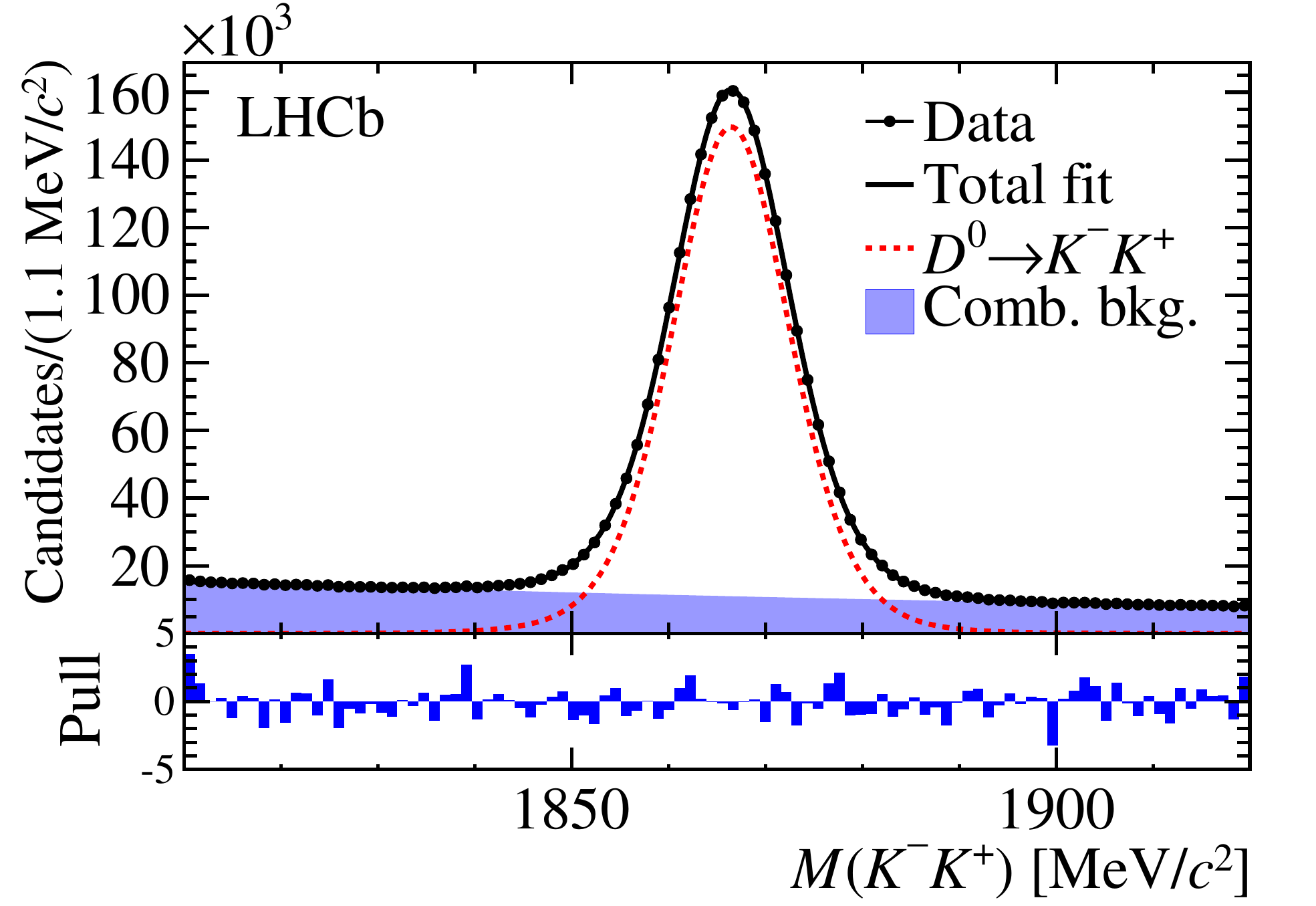}\put(-180,115){(a)}
    \includegraphics[width=0.49\textwidth]{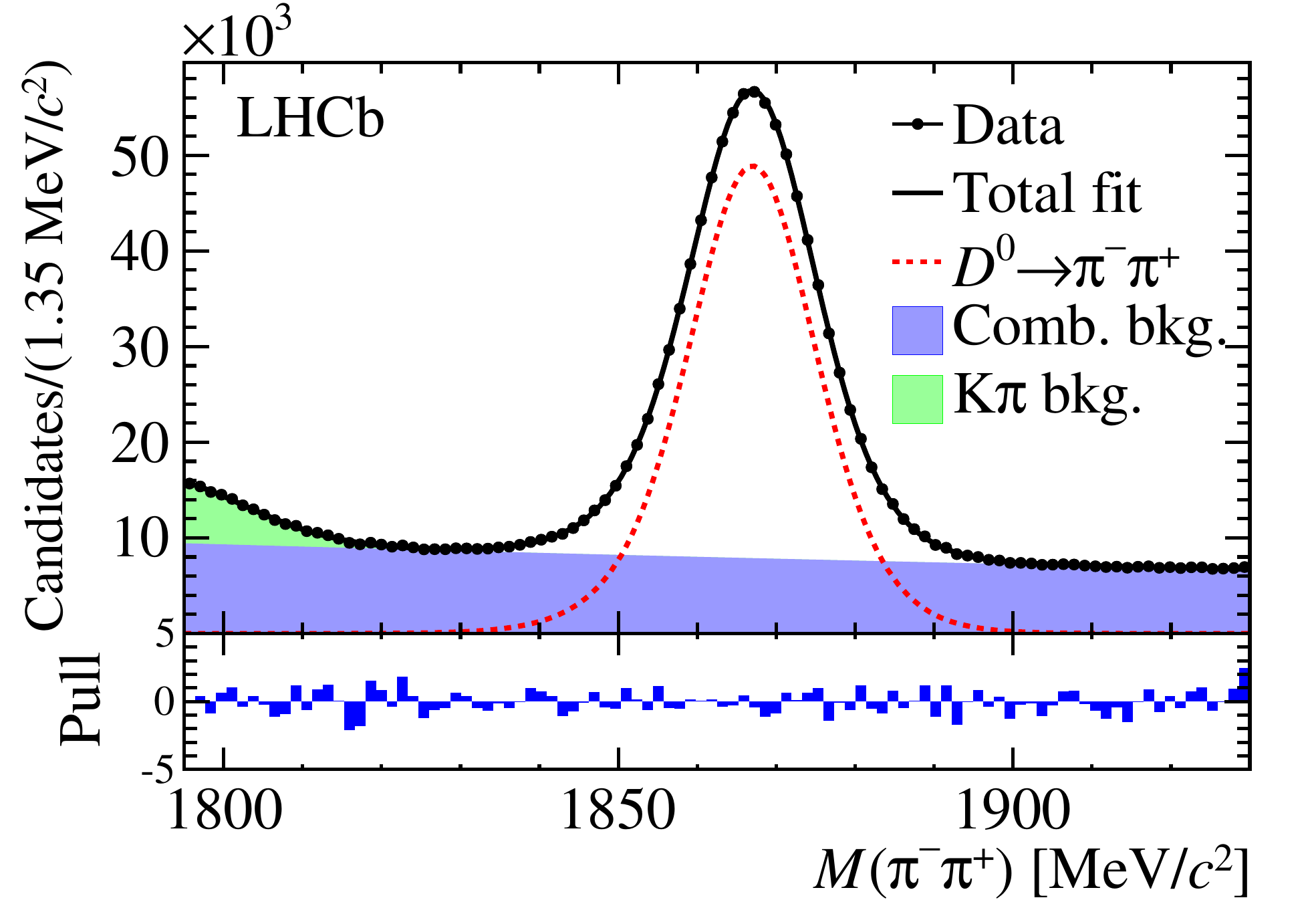}\put(-180,115){(b)}

    \includegraphics[width=0.49\textwidth]{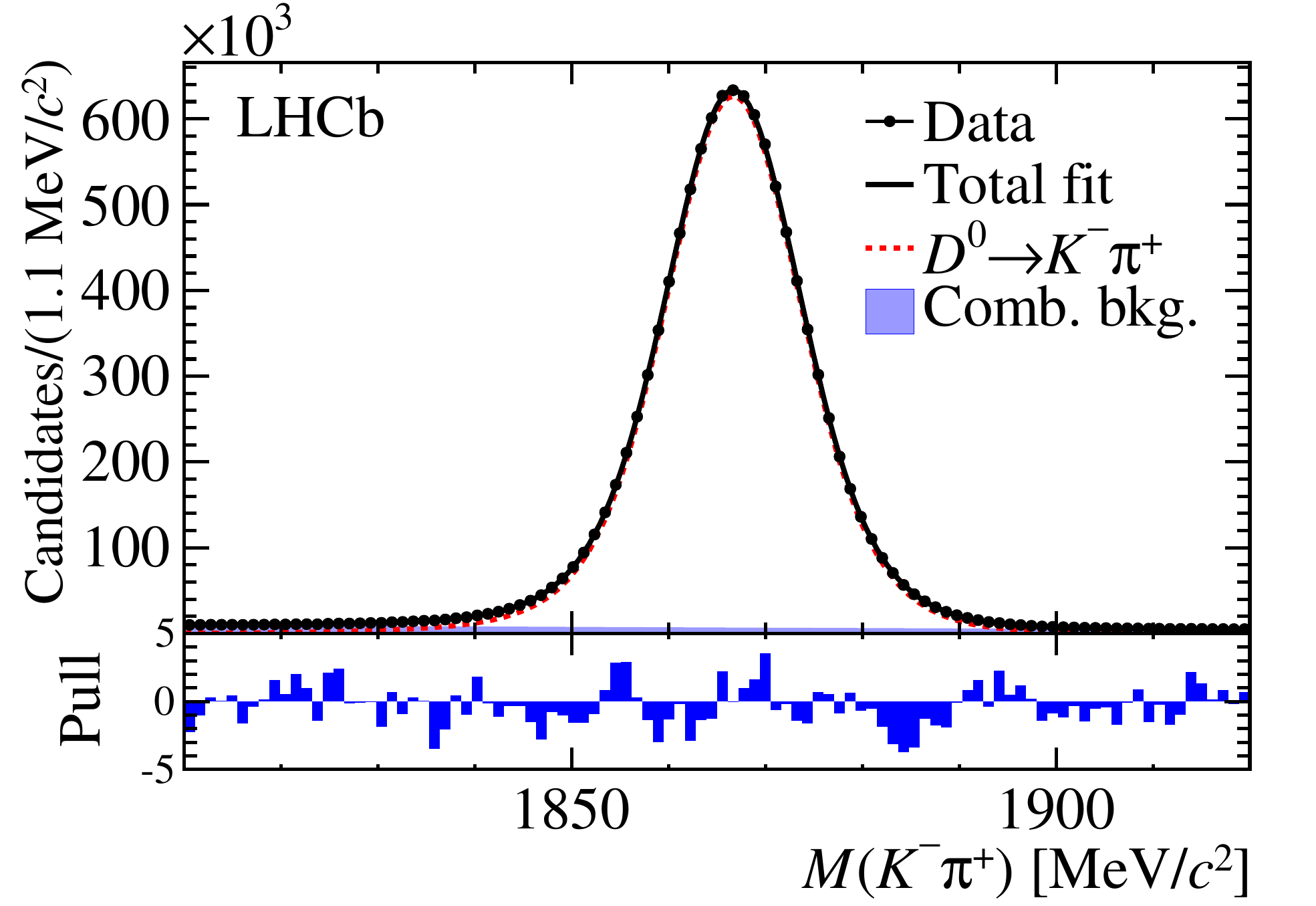}\put(-180,115){(c)}
  \end{center}
  \vspace*{-0.5cm}
  \caption{Invariant mass distributions for (a) \dkk, (b) \dpipi and (c) \dkpi
    candidates. The results of the fits are overlaid. Underneath each plot the
    pull in each mass bin is shown, where the pull is defined as the difference
    between the data point and total fit, divided by the corresponding
    uncertainty.}
  \label{fig:plot_mass}
\end{figure}

% Mass model
The mass distributions for the selected \dkk, \dpipi and \dkpi candidates are
shown in Fig.~\ref{fig:plot_mass}. The numbers of signal candidates are
determined from unbinned extended maximum-likelihood fits in the range 1810 to
1920\mevcc. The signal for all three decay modes is modelled by a sum of three
Gaussian functions. The first two have the same mean, but independent widths;
the third is used to describe a small radiative tail, and has a lower mean and
larger width. The effective width of the signal ranges from $7.1\mevcc$ for \dkk
candidates to $9.3\mevcc$ for \dpipi candidates. As the final states $\Km\Kp$
and $\pim\pip$ are charge symmetric, the shape parameters for the signal are the
same for both \Dz and \Dzb candidates. The combinatorial background is modelled
by an exponential function. In the $\pim\pip$ invariant mass distribution, a
reflection from \dkpi decays is visible in the region below $1820\mevcc$.  This
background component is modelled by a single Gaussian function and the fit range
is extended from 1795 to 1930\mevcc. The shape parameters and overall
normalisation of the background components are allowed to differ between \Dz and
\Dzb candidates. The numbers of signal candidates obtained from these global
fits are $2.34\times10^6$ for \dkk, $0.79\times10^6$ for \dpipi and
$11.31\times10^6$ for \dkpi decays. The latter number corresponds to only half
of the available \dkpi candidates, randomly selected, to reduce the sample size.

% Split in decay time bins
The raw \CP asymmetry is determined from fits to the mass distributions in 50
bins of the \Dz decay time. The fits are performed simultaneously for \Dz and
\Dzb candidates and the asymmetry is determined for each decay-time bin. The
shape parameters and relative normalisation for the third Gaussian function and
for the \dkpi reflection background are fixed from the global fit. All other
parameters are allowed to vary in these fits. In particular, since both the
amount and the composition of background depend on the decay time, the
background parameters are free to vary in each decay-time bin. For decay times
larger than $1\ps$ the relative contribution from combinatorial background
increases. This is due to the exponential decrease of the signal and a less
steep dependence of the combinatorial background on the decay time. The mass
distribution in each decay-time bin is well described by the model.

Events at large \Dz decay times have a larger sensitivity to \AG compared to
events at small decay times, which is balanced by the fewer signal candidates at
large decay times. The binning in \Dz decay time is chosen such that every bin
gives roughly the same statistical contribution to \AG. The value of \AG is
determined from a \chisq fit to the time-dependent asymmetry of
Eq.~(\ref{eq:ACPt}).  The value of \AG and the offset in the asymmetry are
allowed to vary in the fit, while the \Dz lifetime is fixed to $\tau =
410.1\fs$~\cite{PDG2014}. Due to the exponential decay-time distribution, the
average time in each bin is not in the centre of the bin. Therefore, the
background-subtracted~\cite{Pivk:2004ty} average decay time is determined in
each bin and used in the fit. This fit procedure gives unbiased results and
correct uncertainties, as is verified by simulating many experiments with large
samples.

\begin{figure}
  \begin{center}
    \includegraphics[width=0.65\textwidth]{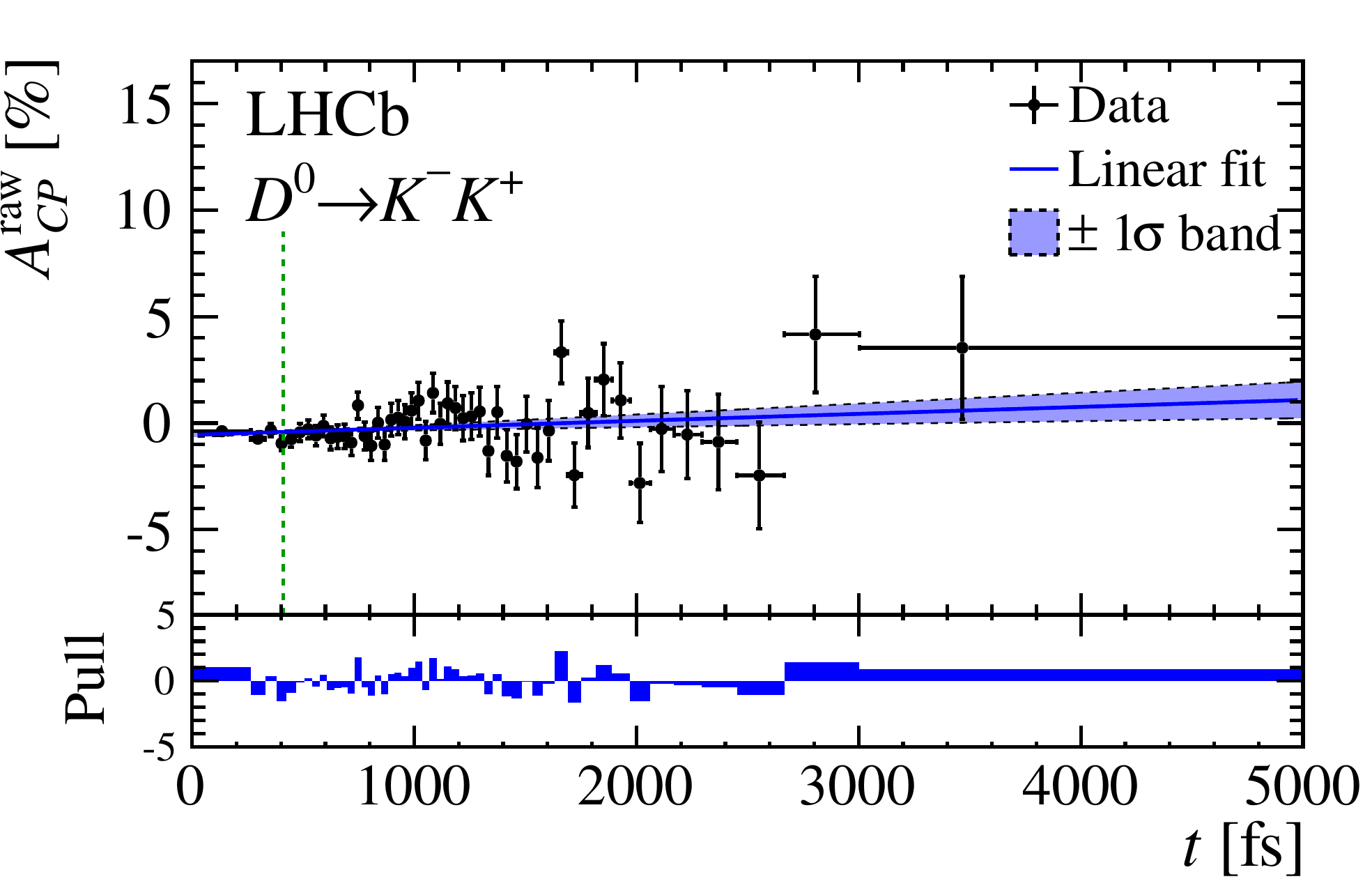}\put(-175,155){\large{(a)}}\\
    \includegraphics[width=0.65\textwidth]{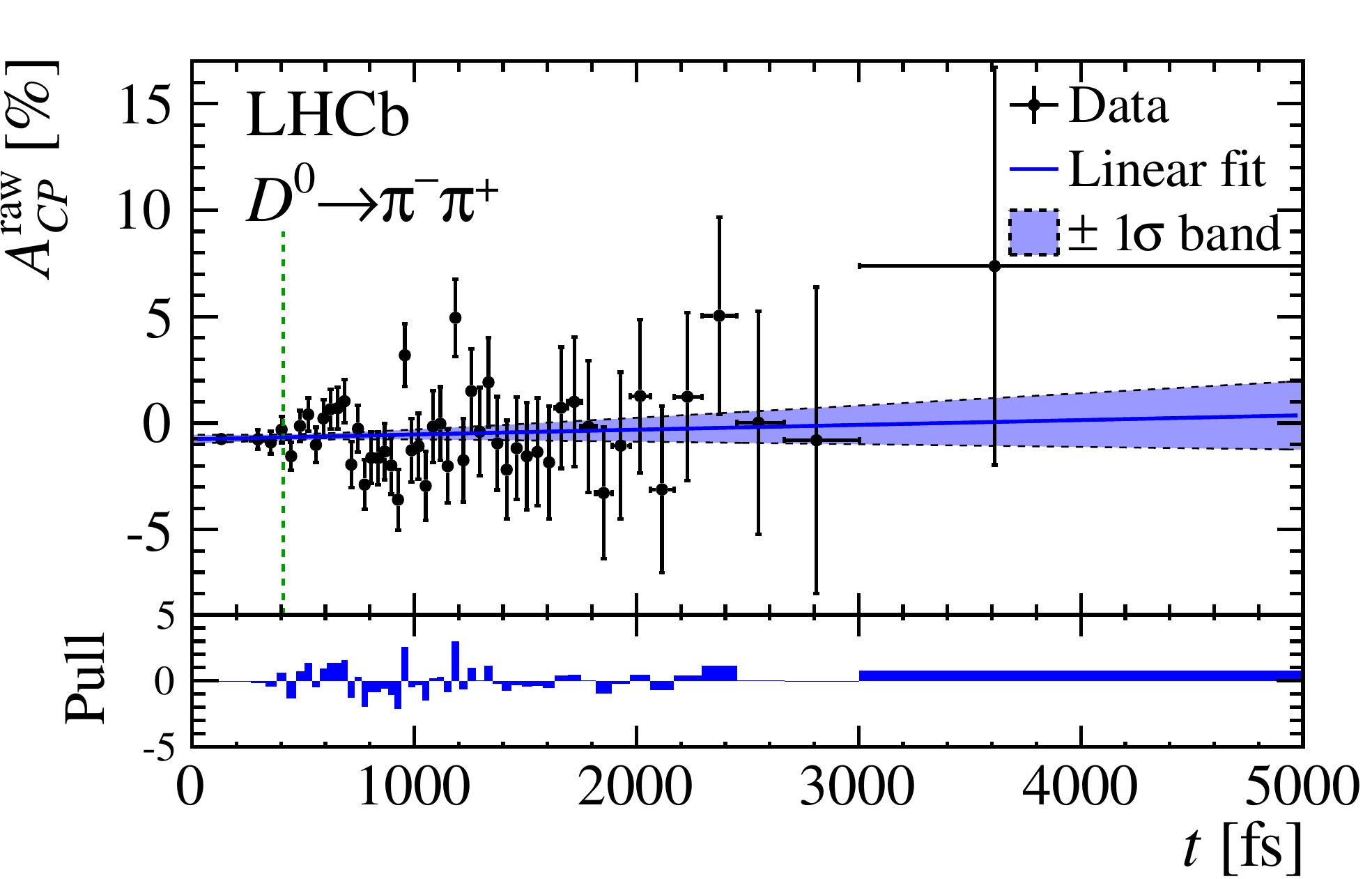}\put(-175,155){\large{(b)}}\\
    \includegraphics[width=0.65\textwidth]{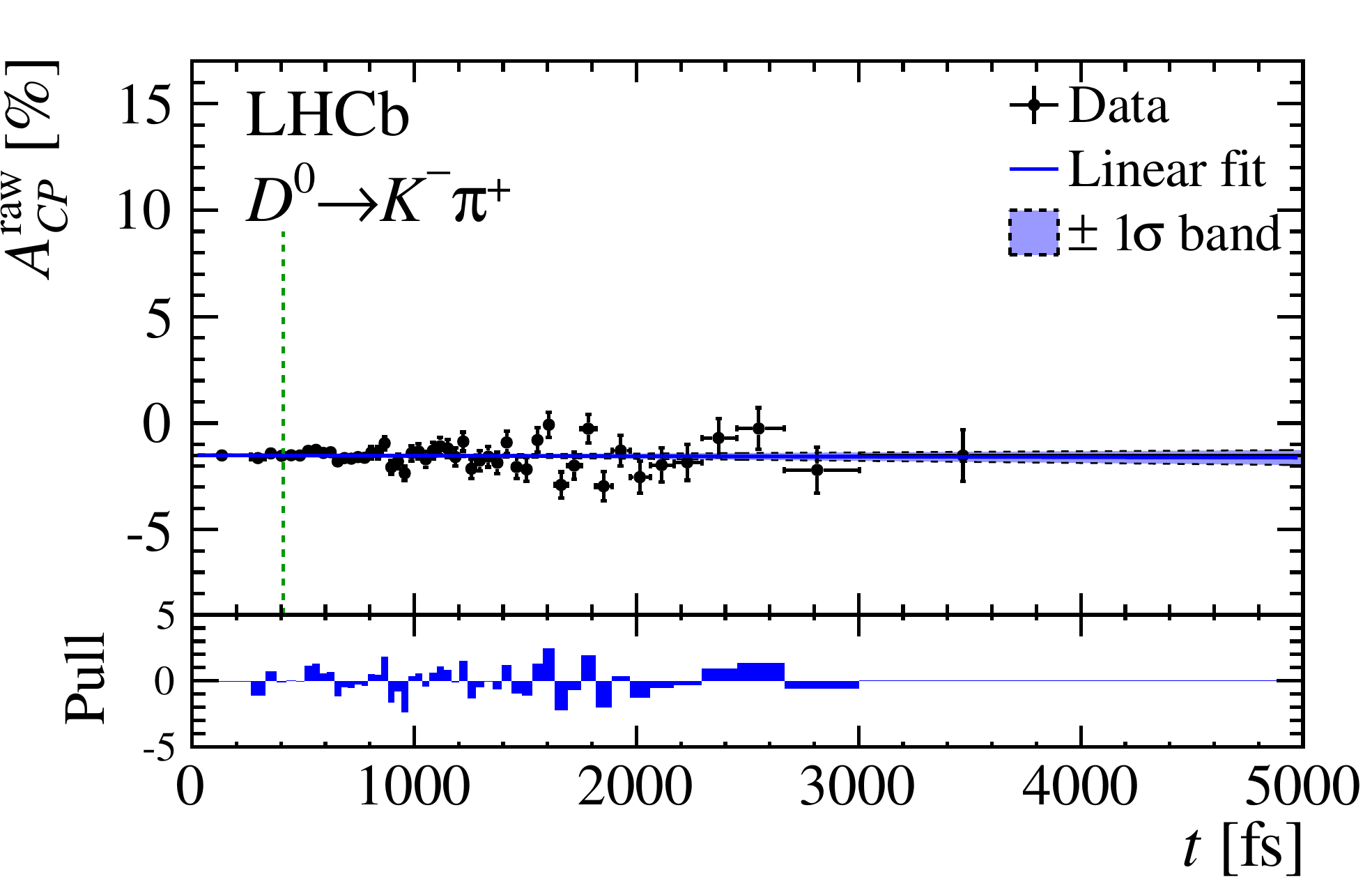}\put(-175,155){\large{(c)}}
  \end{center}
  \vspace*{-0.5cm}
  \caption{Raw \CP asymmetry as function of \Dz decay time for (a) \dkk, (b)
    \dpipi and (c) \dkpi candidates. The results of the \chisq fits are shown as
    blue, solid lines with the $\pm1$ standard-deviation ($\sigma$) bands
    indicated by the dashed lines. The green, dashed lines indicate one \Dz
    lifetime ($\tau=410.1\fs$). Underneath each plot the pull in each time bin
    is shown.}
  \label{fig:plot_ACPt}
\end{figure}

The measured asymmetries in bins of decay time are shown in
Fig.~\ref{fig:plot_ACPt}, including the result of the time-dependent fit. The
results in the three decay channels are
\begin{align*}
  \AGKK   &=  (\AGKKval   \pm \AGKKstat  ) \%  \ , \\
  \AGpipi &=  (\AGpipival \pm \AGpipistat) \%  \ , \\
  \AGKpi  &=  (\phantom{-}\AGKpival  \pm \AGKpistat ) \%  \ ,
\end{align*}
where the uncertainties are statistical only. The values for \AG are compatible
with the assumption of no indirect \CP violation. The fits have good $p$-values
of $54.3\%$ (\dkk), $30.8\%$ (\dpipi) and $14.5\%$ (\dkpi). The measured values
for the raw time-integrated asymmetries, which are sensitive to direct \CP
violation, agree with those reported in Ref.~\cite{LHCb-PAPER-2014-013}.

\section{Systematic uncertainties and consistency checks}
\label{sec:systematics}

\begin{figure}
  \begin{center}
    \includegraphics[width=0.65\textwidth]{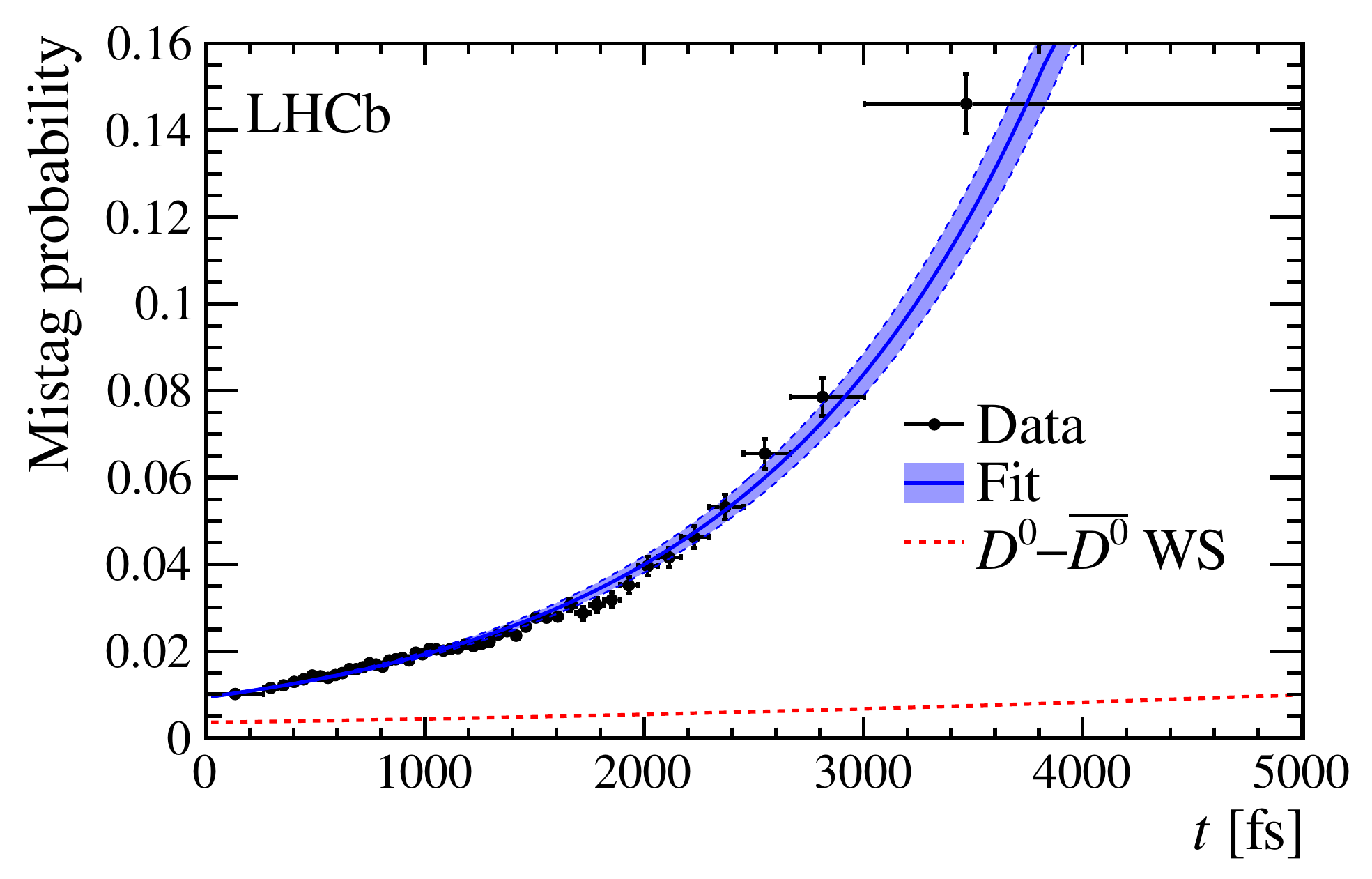}
  \end{center}
  \vspace*{-0.5cm}
  \caption{Mistag probability, before subtracting the contribution from
    wrong-sign (WS) decays, determined with \dkpi candidates. The result of the
    fit to the data points with an exponential function is overlaid (solid, blue
    line). The red, dashed line indicates the expected mistag contribution from
    WS decays.}
  \label{fig:mean_omega}
\end{figure}

% Mistag rate 
The contributions to the systematic uncertainty on \AG are listed in
Table~\ref{tab:syst}. The largest contribution is due to the background coming
from random combinations of muons and \Dz mesons. When the muon has the wrong
charge compared to the real \Dz flavour, this is called a mistag. The mistag
probability ($\mistag$) dilutes the observed asymmetry by a factor
$(1-2\mistag)$. This mistag probability is measured using \dkpi decays,
exploiting the fact that the final state determines the flavour of the \Dz
meson, except for an expected time-dependent wrong-sign fraction due to
\Dz--\Dzb mixing and doubly Cabibbo-suppressed decays. The mistag probability
before correcting for wrong-sign decays is shown in
Fig.~\ref{fig:mean_omega}. After subtracting the (time-dependent) wrong-sign
ratio~\cite{LHCb-PAPER-2013-053}, the mistag probability as function of \Dz
decay time is obtained. The mistag probability is small, with an average around
$1\%$, but it is steeply increasing, reaching 5\% at five \Dz lifetimes. This is
due to the increase of the background fraction from real \Dz mesons from
\bquark-hadron decays combined with a muon from the opposite-side \bquark-hadron
decay. This random-muon background is reconstructed with an apparently longer
lifetime. The time-dependent mistag probability is parameterised by an
exponential function, which is used to determine the shift in \AG. The
systematic uncertainty from this time-dependent mistag probability is $0.006\%$
for the \dkk and $0.008\%$ for the \dpipi decay mode, with a supplementary,
multiplicative scale uncertainty of $0.05$ for both decay modes.

\begin{table}
  \begin{center}
    \caption{Contributions to the systematic uncertainty of \AGKK and
      \AGpipi. The constant and multiplicative scale uncertainties are given
      separately.}
    \label{tab:syst}
    \vspace{0.1cm}
    \begin{tabular}{ l c c c c} \hline
Source of uncertainty & \multicolumn{2}{c}{\dkk} & \multicolumn{2}{c}{\dpipi}\\
                      & constant & scale         & constant & scale \\
\hline
Mistag probability                   & $0.006\%$ & 0.05 & $0.008\%$ & 0.05 \\
Mistag asymmetry                     & $0.016\%$ &      & $0.016\%$ &      \\
Time-dependent efficiency            & $0.010\%$ &      & $0.010\%$ &      \\ 
Detection and production asymmetries & $0.010\%$ &      & $0.010\%$ &      \\
\Dz mass fit model                   & $0.011\%$ &      & $0.007\%$ &      \\
\Dz decay-time resolution            &           & 0.09 &           & 0.07 \\
\Bd--\Bdb mixing                     & $0.007\%$ &      & $0.007\%$ &      \\
\hline 
Quadratic sum      & $0.026\%$ &\AGKKscale & $0.025\%$ & \AGpipiscale \\
\hline
    \end{tabular}
\end{center}
\end{table}

% Mistag asymmetry
The mistag probabilities can potentially differ between positive and negative
muons. Such a mistag asymmetry would give a direct contribution to the observed
asymmetry. The slope of the mistag asymmetry is also obtained from \dkpi
decays. This slope is consistent with no time dependence, and its statistical
uncertainty ($0.016\%$) is included in the systematic uncertainty on \AG.

% Decay time acceptance
The selection of signal candidates, in particular the topological software
trigger, is known to introduce a bias in the observed lifetime. Such a bias
could be charge dependent, thus biasing the measurement of \AG. It is studied
with the \dkpi sample and a sample of $\Dm\to\Kp\pim\pim$ decays from
semileptonic \bquark-hadron decays. No asymmetry of the topological triggers in
single-muon-triggered events is found within an uncertainty of $0.010\%$. This
number is propagated as a systematic uncertainty.

% Detection and production asymmetries
The detection and production asymmetries introduce a constant offset in the raw
time-dependent asymmetries. Since these asymmetries depend on the muon or
\bquark-hadron momentum, they can also introduce a time dependence in case the
momentum spectrum varies between decay-time bins. This effect is tested by
fitting the time-dependent asymmetry after weighting the events so that all
decay-time bins have the same \Dz or muon momentum distribution. The observed
shifts in \AG are within the statistical variations. The shift ($0.010\%$)
observed in the larger \dkpi sample, which has the same production asymmetry and
larger detection asymmetry, is taken as a measure of the systematic uncertainty.

% Mass model
An inaccurate model of the mass distribution can introduce a bias in \AG. The
effect on the observed asymmetries is studied by applying different models in
the fits to the invariant mass distributions. For the signal, a sum of two
Gaussian functions with and without an exponential tail, and for the background
a first and a second-order polynomial are tested. The maximum variation from
the default fit for each decay mode ($0.011\%$ for \dkk; $0.007\%$ for \dpipi)
is taken as a systematic uncertainty on \AG.

% Decay-time resolution
The \Dz decay-time resolution affects the observed time scale, and therefore
changes the measured value of \AG. For each decay mode, the resolution function
is obtained from the simulation, which shows that for the majority of the signal
($90\%$) the decay time is measured with an RMS of about $103\fs$. The remaining
candidates ($10\%$) are measured with an RMS of about $312\fs$. The theoretical
decay rates are convolved with the resolution functions in a large number of
simulated experiments. The effect of the time resolution scales linearly with
the size of \AG. The corresponding scale uncertainty on \AG is $0.09$ for the
\dkk decay mode and $0.07$ for the \dpipi decay mode. Decays where the muon
gives the correct tag but the decay time is biased, \eg, when the muon
originates from a \Ptau lepton in the semileptonic \bquark-hadron decay, are
studied and found to be negligible.

% B0 mixing
About 40\% of the muon-tagged \Dz decays originate from neutral \B
mesons~\cite{LHCb-PAPER-2013-003}. Due to \Bd--\Bdb mixing the observed
production asymmetry depends on the \Bd decay time~\cite{LHCb-PAPER-2014-053}. A
correlation between the \Bd and \Dz decay times may result in a shift in the
measured value of \AG.  The effect of this correlation, determined from
simulation, together with a $1\%$ \Bd production
asymmetry~\cite{LHCb-PAPER-2014-042, LHCb-PAPER-2014-053}, is estimated to be a
shift of $0.007\%$ in the observed value of \AG. This is taken as systematic
uncertainty.

% Systematic overview
Possible shifts in \AG coming from the $1.5\fs$ uncertainty on the world-average
\Dz lifetime~\cite{PDG2014}, from the uncertainty on the momentum scale and
detector length scale~\cite{LHCb-DP-2014-001,LHCb-PAPER-2013-006} and from
potential biases in the fit method are negligible.

The scale uncertainty (cf. Table~\ref{tab:syst}) gives a small contribution to
the overall systematic uncertainty, which depends on the true value of \AG. In
order to present a single systematic uncertainty, the effect of the scale
uncertainty is evaluated with a Neyman construction~\cite{Neyman}. For each
true value of \AG, the absolute size of the scale uncertainty is known and added
in quadrature to the constant uncertainty. In this way, a confidence belt of
observed values versus true values is constructed. This procedure gives a
slightly asymmetric systematic uncertainty, which is
$^{+\AGKKsystP}_{-\AGKKsystM}\%$ for the \dkk decay channel and
$^{+\AGpipisystP}_{-\AGpipisystM}\%$ for the \dpipi decay channel. Except for
the contribution from the mass fit model, all contributions to the systematic
uncertainty are fully correlated, resulting in an overall correlation
coefficient of $89\%$ between the systematic uncertainties of \AGKK and \AGpipi.

\begin{figure}
  \begin{center}
    \includegraphics[width=0.49\textwidth]{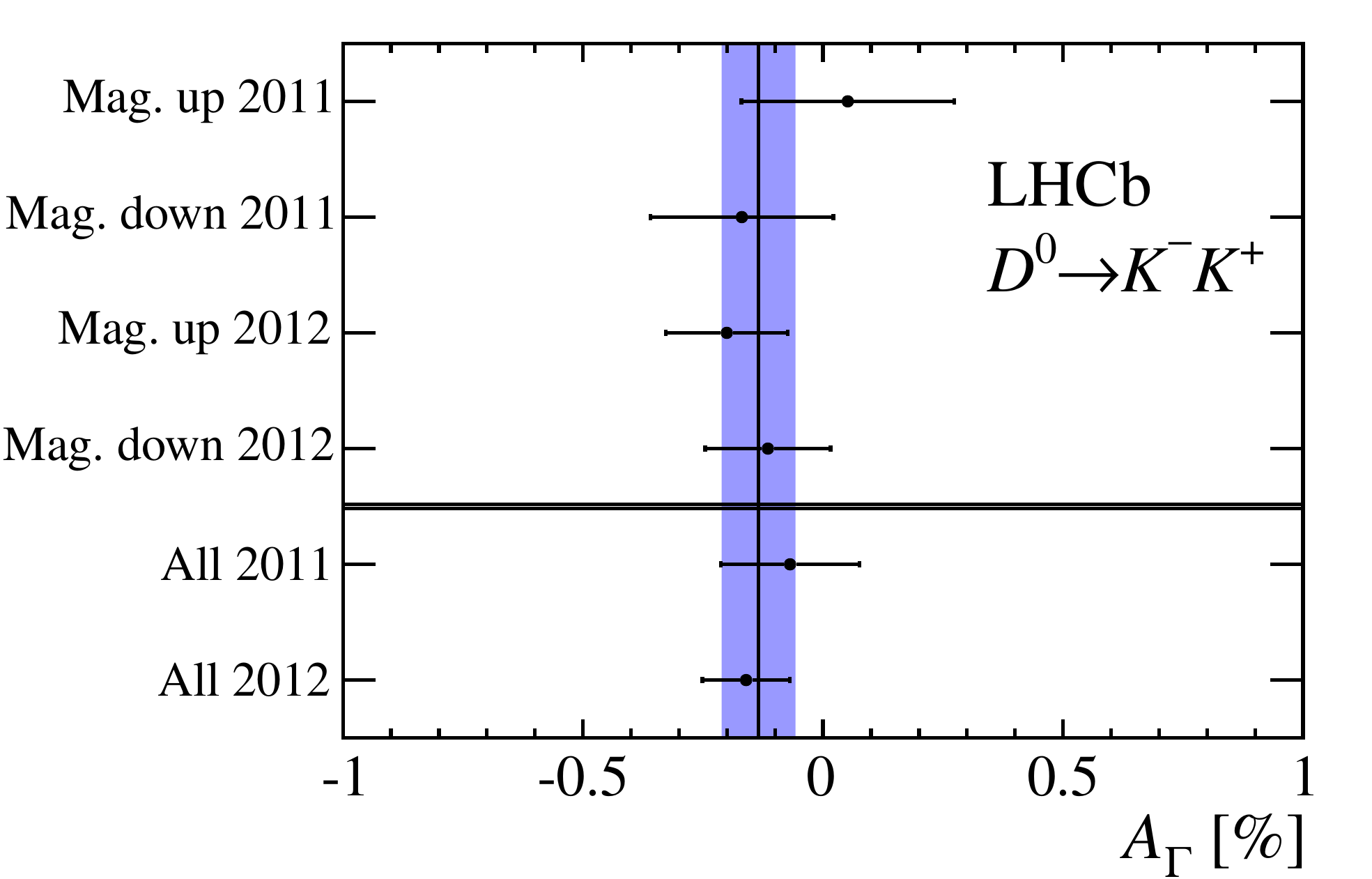}\put(-160,120){(a)}
    \includegraphics[width=0.49\textwidth]{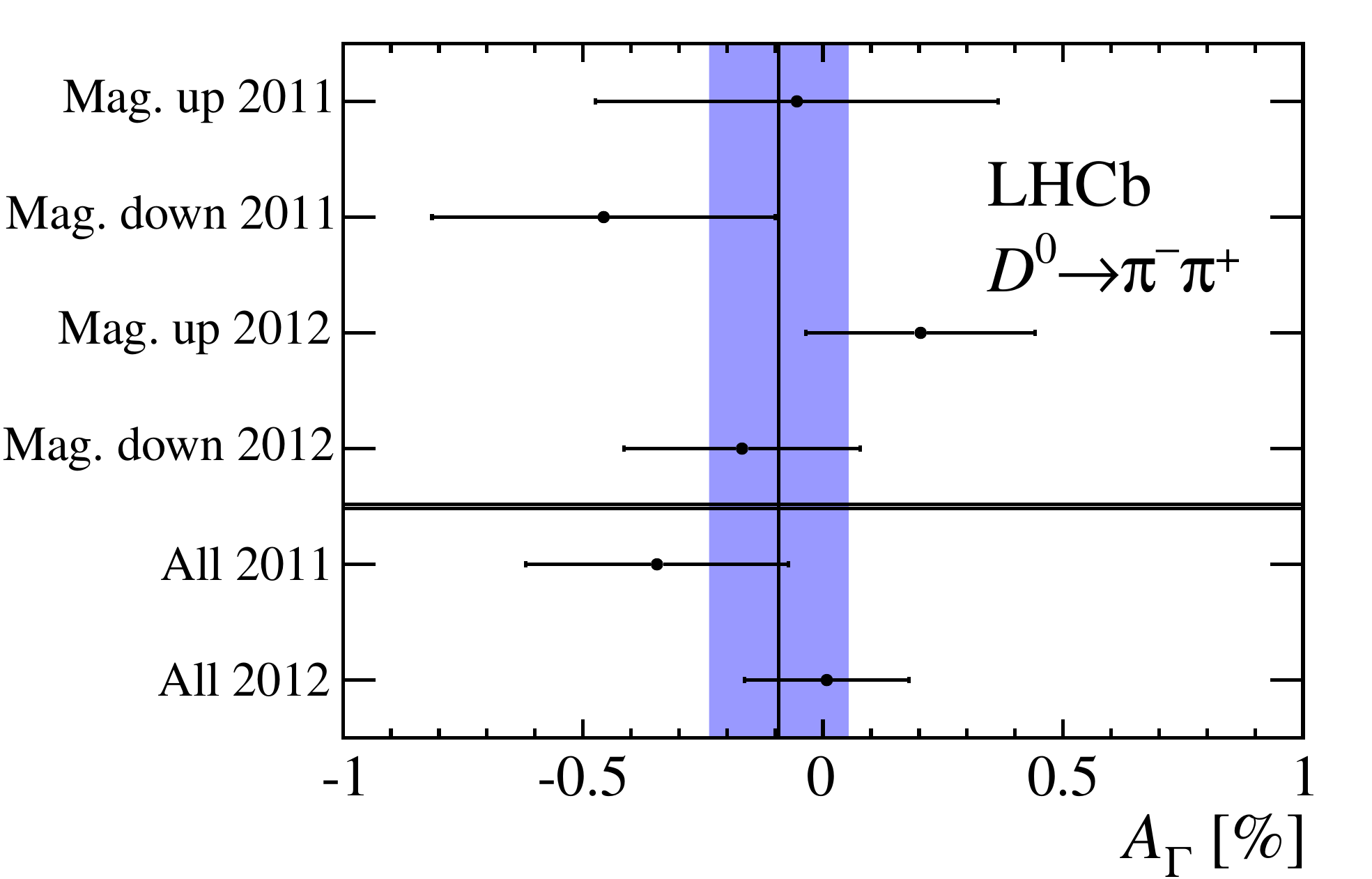}\put(-160,120){(b)}

    \includegraphics[width=0.49\textwidth]{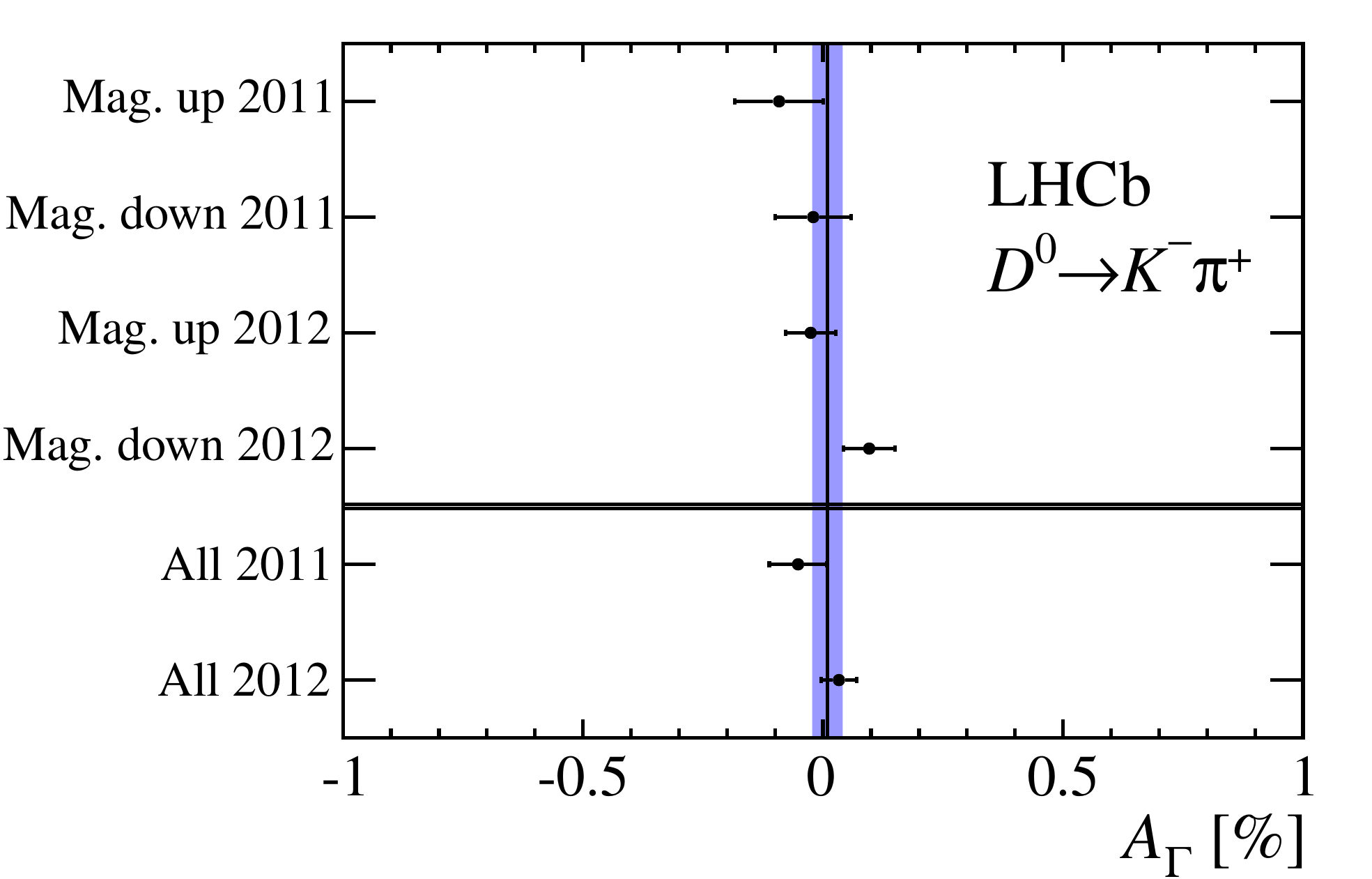}\put(-160,120){(c)}
  \end{center}
  \vspace*{-0.5cm}
  \caption{Measured values of \AG for different magnet polarities and
    data-taking periods for (a) \dkk, (b) \dpipi and (c) \dkpi decays. The
    vertical line and error band indicate the average \AG obtained from the
    combined data set. The error bars indicate the statistical uncertainty
    only.}
  \label{fig:datasplit}
\end{figure}

Additional checks have been performed to determine potential sensitivity of the
measurements on the data-taking conditions, detector configuration, and analysis
procedure. Changing to a finer decay-time binning yields compatible
results. Potential effects on the measurement of \AG coming from detection
asymmetries are expected to appear when dividing the data set by magnet polarity
and data-taking period. Detection asymmetries originating from a left-right
asymmetric detector change sign when reversing the magnet polarity. Similarly,
during the two data-taking periods, detection asymmetries and production
asymmetries might have changed due to different running conditions. As shown in
Fig.~\ref{fig:datasplit}, there is no significant variation of \AG across
various configurations. Also splitting the data set according to the number of
primary vertices or in bins of the \B decay time does not show any deviation in
the measured values of
\AG.

\section{Conclusions}
\label{sec:conclusion}

The time-dependent \CP asymmetries in \dkk and \dpipi decays are measured using
muon-tagged \Dz mesons originating from semileptonic \bquark-hadron decays in
the 3.0\invfb data set collected with the \lhcb detector in 2011 and 2012. The
asymmetries in the effective lifetimes are measured to be
\begin{align*}
  \AGKK  = (\AGKKval \pm \AGKKstat \; {}^{+\AGKKsystP}_{-\AGKKsystM})\% \ , \\
  \AGpipi = (\AGpipival\pm \AGpipistat \; {}^{+\AGpipisystP}_{-\AGpipisystM})\% \ ,
\end{align*}
where the first uncertainty is statistical and the second systematic.  Assuming
that indirect \CP violation in \Dz decays is universal~\cite{Grossman:2006jg},
and accounting for the correlation in the systematic uncertainties, the average
of the two measurements becomes $\AG = (-0.125 \pm 0.073)\%$. The results in
this paper are uncorrelated with the time-integrated asymmetries reported in
Ref.~\cite{LHCb-PAPER-2014-013}. The results are consistent with other \AG
measurements~\cite{Lees:2012qh, LHCb-PAPER-2013-054, Aaltonen:2014efa}, and
independent of the \AG measurements~\cite{LHCb-PAPER-2013-054} from \lhcb using
\Dz mesons from $\Dstarp\to\Dz\pip$ decays. They are consistent with the
hypothesis of no indirect \CP violation in \dkk and \dpipi decays.

% Do not include this in analysis note and conference reports
\section*{Acknowledgements}
 
\noindent We express our gratitude to our colleagues in the CERN
accelerator departments for the excellent performance of the LHC. We
thank the technical and administrative staff at the LHCb
institutes. We acknowledge support from CERN and from the national
agencies: CAPES, CNPq, FAPERJ and FINEP (Brazil); NSFC (China);
CNRS/IN2P3 (France); BMBF, DFG, HGF and MPG (Germany); INFN (Italy); 
FOM and NWO (The Netherlands); MNiSW and NCN (Poland); MEN/IFA (Romania); 
MinES and FANO (Russia); MinECo (Spain); SNSF and SER (Switzerland); 
NASU (Ukraine); STFC (United Kingdom); NSF (USA).
The Tier1 computing centres are supported by IN2P3 (France), KIT and BMBF 
(Germany), INFN (Italy), NWO and SURF (The Netherlands), PIC (Spain), GridPP 
(United Kingdom).
We are indebted to the communities behind the multiple open 
source software packages on which we depend. We are also thankful for the 
computing resources and the access to software R\&D tools provided by Yandex LLC (Russia).
Individual groups or members have received support from 
EPLANET, Marie Sk\l{}odowska-Curie Actions and ERC (European Union), 
Conseil g\'{e}n\'{e}ral de Haute-Savoie, Labex ENIGMASS and OCEVU, 
R\'{e}gion Auvergne (France), RFBR (Russia), XuntaGal and GENCAT (Spain), Royal Society and Royal
Commission for the Exhibition of 1851 (United Kingdom).

\addcontentsline{toc}{section}{References}
\setboolean{inbibliography}{true}
\bibliographystyle{LHCb}
\bibliography{main,LHCb-PAPER,LHCb-DP}

\ifx\mcitethebibliography\mciteundefinedmacro
\PackageError{LHCb.bst}{mciteplus.sty has not been loaded}
{This bibstyle requires the use of the mciteplus package.}\fi
\providecommand{\href}[2]{#2}
\begin{mcitethebibliography}{10}
\mciteSetBstSublistMode{n}
\mciteSetBstMaxWidthForm{subitem}{\alph{mcitesubitemcount})}
\mciteSetBstSublistLabelBeginEnd{\mcitemaxwidthsubitemform\space}
{\relax}{\relax}

\bibitem{Lees:2012qh}
BaBar collaboration, J.~P. Lees {\em et~al.},
  \ifthenelse{\boolean{articletitles}}{\emph{{Measurement of \Dz--\Dzb mixing
  and \CP violation in two-body \Dz decays}},
  }{}\href{http://dx.doi.org/10.1103/PhysRevD.87.012004}{Phys.\ Rev.\
  \textbf{D87} (2013) 012004}, \href{http://arxiv.org/abs/1209.3896}{{\tt
  arXiv:1209.3896}}\relax
\mciteBstWouldAddEndPuncttrue
\mciteSetBstMidEndSepPunct{\mcitedefaultmidpunct}
{\mcitedefaultendpunct}{\mcitedefaultseppunct}\relax
\EndOfBibitem
\bibitem{delAmoSanchez:2010xz}
BaBar collaboration, P.~del Amo~Sanchez {\em et~al.},
  \ifthenelse{\boolean{articletitles}}{\emph{{Measurement of $\Dz$--$\Dzb$
  mixing parameters using $\Dz\to\KS\pip\pim$ and $\Dz\to\KS\Kp\Km$ decays}},
  }{}\href{http://dx.doi.org/10.1103/PhysRevLett.105.081803}{Phys.\ Rev.\
  Lett.\  \textbf{105} (2010) 081803},
  \href{http://arxiv.org/abs/1004.5053}{{\tt arXiv:1004.5053}}\relax
\mciteBstWouldAddEndPuncttrue
\mciteSetBstMidEndSepPunct{\mcitedefaultmidpunct}
{\mcitedefaultendpunct}{\mcitedefaultseppunct}\relax
\EndOfBibitem
\bibitem{LHCb-PAPER-2013-053}
LHCb collaboration, R.~Aaij {\em et~al.},
  \ifthenelse{\boolean{articletitles}}{\emph{{Measurement of \Dz--\Dzb mixing
  parameters and search for \CP violation using $D^0\to K^+\pi^-$ decays}},
  }{}\href{http://dx.doi.org/10.1103/PhysRevLett.111.251801}{Phys.\ Rev.\
  Lett.\  \textbf{111} (2013) 251801},
  \href{http://arxiv.org/abs/1309.6534}{{\tt arXiv:1309.6534}}\relax
\mciteBstWouldAddEndPuncttrue
\mciteSetBstMidEndSepPunct{\mcitedefaultmidpunct}
{\mcitedefaultendpunct}{\mcitedefaultseppunct}\relax
\EndOfBibitem
\bibitem{Aaltonen:2013pja}
CDF collaboration, T.~A. Aaltonen {\em et~al.},
  \ifthenelse{\boolean{articletitles}}{\emph{{Observation of \Dz--\Dzb mixing
  using the CDF II detector}},
  }{}\href{http://dx.doi.org/10.1103/PhysRevLett.111.231802}{Phys.\ Rev.\
  Lett.\  \textbf{111} (2013) 231802},
  \href{http://arxiv.org/abs/1309.4078}{{\tt arXiv:1309.4078}}\relax
\mciteBstWouldAddEndPuncttrue
\mciteSetBstMidEndSepPunct{\mcitedefaultmidpunct}
{\mcitedefaultendpunct}{\mcitedefaultseppunct}\relax
\EndOfBibitem
\bibitem{Ko:2014qvu}
Belle collaboration, B.~R. Ko {\em et~al.},
  \ifthenelse{\boolean{articletitles}}{\emph{{Observation of $\Dz$--$\Dzb$
  mixing in $e^+e^-$ collisions}},
  }{}\href{http://dx.doi.org/10.1103/PhysRevLett.112.111801,
  10.1103/PhysRevLett.112.139903}{Phys.\ Rev.\ Lett.\  \textbf{112} (2014)
  111801}, \href{http://arxiv.org/abs/1401.3402}{{\tt arXiv:1401.3402}}\relax
\mciteBstWouldAddEndPuncttrue
\mciteSetBstMidEndSepPunct{\mcitedefaultmidpunct}
{\mcitedefaultendpunct}{\mcitedefaultseppunct}\relax
\EndOfBibitem
\bibitem{Peng:2014oda}
Belle collaboration, T.~Peng {\em et~al.},
  \ifthenelse{\boolean{articletitles}}{\emph{{Measurement of $\Dz$--$\Dzb$
  mixing and search for indirect CP violation using $\Dz\to\KS\pip\pim$
  decays}}, }{}\href{http://dx.doi.org/10.1103/PhysRevD.89.091103}{Phys.\ Rev.\
   \textbf{D89} (2014) 091103}, \href{http://arxiv.org/abs/1404.2412}{{\tt
  arXiv:1404.2412}}\relax
\mciteBstWouldAddEndPuncttrue
\mciteSetBstMidEndSepPunct{\mcitedefaultmidpunct}
{\mcitedefaultendpunct}{\mcitedefaultseppunct}\relax
\EndOfBibitem
\bibitem{HFAG}
Heavy Flavor Averaging Group, Y.~Amhis {\em et~al.},
  \ifthenelse{\boolean{articletitles}}{\emph{{Averages of $b$-hadron,
  $c$-hadron, and $\tau$-lepton properties as of summer 2014}},
  }{}\href{http://arxiv.org/abs/1412.7515}{{\tt arXiv:1412.7515}}, {updated
  results and plots available at
  \href{http://www.slac.stanford.edu/xorg/hfag/}{{\tt
  http://www.slac.stanford.edu/xorg/hfag/}}}\relax
\mciteBstWouldAddEndPuncttrue
\mciteSetBstMidEndSepPunct{\mcitedefaultmidpunct}
{\mcitedefaultendpunct}{\mcitedefaultseppunct}\relax
\EndOfBibitem
\bibitem{Bianco:2003vb}
S.~Bianco, F.~L. Fabbri, D.~Benson, and I.~Bigi,
  \ifthenelse{\boolean{articletitles}}{\emph{{A Cicerone for the physics of
  charm}}, }{}\href{http://dx.doi.org/10.1393/ncr/i2003-10003-1}{Riv.\ Nuovo
  Cim.\  \textbf{26N7} (2003) 1},
  \href{http://arxiv.org/abs/hep-ex/0309021}{{\tt arXiv:hep-ex/0309021}}\relax
\mciteBstWouldAddEndPuncttrue
\mciteSetBstMidEndSepPunct{\mcitedefaultmidpunct}
{\mcitedefaultendpunct}{\mcitedefaultseppunct}\relax
\EndOfBibitem
\bibitem{Bobrowski:2010xg}
M.~Bobrowski, A.~Lenz, J.~Riedl, and J.~Rohrwild,
  \ifthenelse{\boolean{articletitles}}{\emph{{How large can the SM contribution
  to \CP violation in $\Dz$--$\Dzb$ mixing be?}},
  }{}\href{http://dx.doi.org/10.1007/JHEP03(2010)009}{JHEP \textbf{03} (2010)
  009}, \href{http://arxiv.org/abs/1002.4794}{{\tt arXiv:1002.4794}}\relax
\mciteBstWouldAddEndPuncttrue
\mciteSetBstMidEndSepPunct{\mcitedefaultmidpunct}
{\mcitedefaultendpunct}{\mcitedefaultseppunct}\relax
\EndOfBibitem
\bibitem{Grossman:2006jg}
Y.~Grossman, A.~L. Kagan, and Y.~Nir,
  \ifthenelse{\boolean{articletitles}}{\emph{{New physics and \CP violation in
  singly Cabibbo suppressed \D decays}},
  }{}\href{http://dx.doi.org/10.1103/PhysRevD.75.036008}{Phys.\ Rev.\
  \textbf{D75} (2007) 036008}, \href{http://arxiv.org/abs/hep-ph/0609178}{{\tt
  arXiv:hep-ph/0609178}}\relax
\mciteBstWouldAddEndPuncttrue
\mciteSetBstMidEndSepPunct{\mcitedefaultmidpunct}
{\mcitedefaultendpunct}{\mcitedefaultseppunct}\relax
\EndOfBibitem
\bibitem{LHCb-PAPER-2013-054}
LHCb collaboration, R.~Aaij {\em et~al.},
  \ifthenelse{\boolean{articletitles}}{\emph{{Measurements of indirect \CP
  asymmetries in $D^0\to K^-K^+$ and $D^0\to\pi^-\pi^+$ decays}},
  }{}\href{http://dx.doi.org/10.1103/PhysRevLett.112.041801}{Phys.\ Rev.\
  Lett.\  \textbf{112} (2014) 041801},
  \href{http://arxiv.org/abs/1310.7201}{{\tt arXiv:1310.7201}}\relax
\mciteBstWouldAddEndPuncttrue
\mciteSetBstMidEndSepPunct{\mcitedefaultmidpunct}
{\mcitedefaultendpunct}{\mcitedefaultseppunct}\relax
\EndOfBibitem
\bibitem{Aaltonen:2014efa}
CDF collaboration, T.~A. Aaltonen {\em et~al.},
  \ifthenelse{\boolean{articletitles}}{\emph{{Measurement of indirect
  \CP-violating asymmetries in $\Dz\to\Kp\Km$ and $\Dz\to\pip\pim$ decays at
  CDF}}, }{}\href{http://dx.doi.org/10.1103/PhysRevD.90.111103}{Phys.\ Rev.\
  \textbf{D90} (2014) 111103}, \href{http://arxiv.org/abs/1410.5435}{{\tt
  arXiv:1410.5435}}\relax
\mciteBstWouldAddEndPuncttrue
\mciteSetBstMidEndSepPunct{\mcitedefaultmidpunct}
{\mcitedefaultendpunct}{\mcitedefaultseppunct}\relax
\EndOfBibitem
\bibitem{LHCb-PAPER-2014-013}
LHCb collaboration, R.~Aaij {\em et~al.},
  \ifthenelse{\boolean{articletitles}}{\emph{{Measurement of \CP asymmetry in
  $D^0 \to K^- K^+$ and $D^0 \to \pi^- \pi^+$ decays}},
  }{}\href{http://dx.doi.org/10.1007/JHEP07(2014)041}{JHEP \textbf{07} (2014)
  041}, \href{http://arxiv.org/abs/1405.2797}{{\tt arXiv:1405.2797}}\relax
\mciteBstWouldAddEndPuncttrue
\mciteSetBstMidEndSepPunct{\mcitedefaultmidpunct}
{\mcitedefaultendpunct}{\mcitedefaultseppunct}\relax
\EndOfBibitem
\bibitem{Aaltonen:2011se}
CDF collaboration, T.~A. Aaltonen {\em et~al.},
  \ifthenelse{\boolean{articletitles}}{\emph{{Measurement of \CP-violating
  asymmetries in $\Dz\to\pip\pim$ and $\Dz\to\Kp\Km$ decays at CDF}},
  }{}\href{http://dx.doi.org/10.1103/PhysRevD.85.012009}{Phys.\ Rev.\
  \textbf{D85} (2012) 012009}, \href{http://arxiv.org/abs/1111.5023}{{\tt
  arXiv:1111.5023}}\relax
\mciteBstWouldAddEndPuncttrue
\mciteSetBstMidEndSepPunct{\mcitedefaultmidpunct}
{\mcitedefaultendpunct}{\mcitedefaultseppunct}\relax
\EndOfBibitem
\bibitem{Gersabeck:2011xj}
M.~Gersabeck {\em et~al.}, \ifthenelse{\boolean{articletitles}}{\emph{{On the
  interplay of direct and indirect \CP violation in the charm sector}},
  }{}\href{http://dx.doi.org/10.1088/0954-3899/39/4/045005}{J.\ Phys.\
  \textbf{G39} (2012) 045005}, \href{http://arxiv.org/abs/1111.6515}{{\tt
  arXiv:1111.6515}}\relax
\mciteBstWouldAddEndPuncttrue
\mciteSetBstMidEndSepPunct{\mcitedefaultmidpunct}
{\mcitedefaultendpunct}{\mcitedefaultseppunct}\relax
\EndOfBibitem
\bibitem{Kagan:2009gb}
A.~L. Kagan and M.~D. Sokoloff,
  \ifthenelse{\boolean{articletitles}}{\emph{{Indirect \CP violation and
  implications for $\Dz$--$\Dzb$ and $B_s$--$\Bbar{}_s$ mixing}},
  }{}\href{http://dx.doi.org/10.1103/PhysRevD.80.076008}{Phys.\ Rev.\
  \textbf{D80} (2009) 076008}, \href{http://arxiv.org/abs/0907.3917}{{\tt
  arXiv:0907.3917}}\relax
\mciteBstWouldAddEndPuncttrue
\mciteSetBstMidEndSepPunct{\mcitedefaultmidpunct}
{\mcitedefaultendpunct}{\mcitedefaultseppunct}\relax
\EndOfBibitem
\bibitem{Alves:2008zz}
LHCb collaboration, A.~A. Alves~Jr.\ {\em et~al.},
  \ifthenelse{\boolean{articletitles}}{\emph{{The \lhcb detector at the LHC}},
  }{}\href{http://dx.doi.org/10.1088/1748-0221/3/08/S08005}{JINST \textbf{3}
  (2008) S08005}\relax
\mciteBstWouldAddEndPuncttrue
\mciteSetBstMidEndSepPunct{\mcitedefaultmidpunct}
{\mcitedefaultendpunct}{\mcitedefaultseppunct}\relax
\EndOfBibitem
\bibitem{LHCb-DP-2014-002}
LHCb collaboration, R.~Aaij {\em et~al.},
  \ifthenelse{\boolean{articletitles}}{\emph{{LHCb detector performance}},
  }{}\href{http://arxiv.org/abs/1412.6352}{{\tt arXiv:1412.6352}}\relax
\mciteBstWouldAddEndPuncttrue
\mciteSetBstMidEndSepPunct{\mcitedefaultmidpunct}
{\mcitedefaultendpunct}{\mcitedefaultseppunct}\relax
\EndOfBibitem
\bibitem{LHCb-DP-2012-004}
R.~Aaij {\em et~al.}, \ifthenelse{\boolean{articletitles}}{\emph{{The \lhcb
  trigger and its performance in 2011}},
  }{}\href{http://dx.doi.org/10.1088/1748-0221/8/04/P04022}{JINST \textbf{8}
  (2013) P04022}, \href{http://arxiv.org/abs/1211.3055}{{\tt
  arXiv:1211.3055}}\relax
\mciteBstWouldAddEndPuncttrue
\mciteSetBstMidEndSepPunct{\mcitedefaultmidpunct}
{\mcitedefaultendpunct}{\mcitedefaultseppunct}\relax
\EndOfBibitem
\bibitem{Sjostrand:2006za}
T.~Sj\"{o}strand, S.~Mrenna, and P.~Skands,
  \ifthenelse{\boolean{articletitles}}{\emph{{PYTHIA 6.4 physics and manual}},
  }{}\href{http://dx.doi.org/10.1088/1126-6708/2006/05/026}{JHEP \textbf{05}
  (2006) 026}, \href{http://arxiv.org/abs/hep-ph/0603175}{{\tt
  arXiv:hep-ph/0603175}}\relax
\mciteBstWouldAddEndPuncttrue
\mciteSetBstMidEndSepPunct{\mcitedefaultmidpunct}
{\mcitedefaultendpunct}{\mcitedefaultseppunct}\relax
\EndOfBibitem
\bibitem{Sjostrand:2007gs}
T.~Sj\"{o}strand, S.~Mrenna, and P.~Skands,
  \ifthenelse{\boolean{articletitles}}{\emph{{A brief introduction to PYTHIA
  8.1}}, }{}\href{http://dx.doi.org/10.1016/j.cpc.2008.01.036}{Comput.\ Phys.\
  Commun.\  \textbf{178} (2008) 852},
  \href{http://arxiv.org/abs/0710.3820}{{\tt arXiv:0710.3820}}\relax
\mciteBstWouldAddEndPuncttrue
\mciteSetBstMidEndSepPunct{\mcitedefaultmidpunct}
{\mcitedefaultendpunct}{\mcitedefaultseppunct}\relax
\EndOfBibitem
\bibitem{LHCb-PROC-2010-056}
I.~Belyaev {\em et~al.}, \ifthenelse{\boolean{articletitles}}{\emph{{Handling
  of the generation of primary events in Gauss, the LHCb simulation
  framework}}, }{}\href{http://dx.doi.org/10.1109/NSSMIC.2010.5873949}{Nuclear
  Science Symposium Conference Record (NSS/MIC) \textbf{IEEE} (2010)
  1155}\relax
\mciteBstWouldAddEndPuncttrue
\mciteSetBstMidEndSepPunct{\mcitedefaultmidpunct}
{\mcitedefaultendpunct}{\mcitedefaultseppunct}\relax
\EndOfBibitem
\bibitem{Lange:2001uf}
D.~J. Lange, \ifthenelse{\boolean{articletitles}}{\emph{{The EvtGen particle
  decay simulation package}},
  }{}\href{http://dx.doi.org/10.1016/S0168-9002(01)00089-4}{Nucl.\ Instrum.\
  Meth.\  \textbf{A462} (2001) 152}\relax
\mciteBstWouldAddEndPuncttrue
\mciteSetBstMidEndSepPunct{\mcitedefaultmidpunct}
{\mcitedefaultendpunct}{\mcitedefaultseppunct}\relax
\EndOfBibitem
\bibitem{Golonka:2005pn}
P.~Golonka and Z.~Was, \ifthenelse{\boolean{articletitles}}{\emph{{PHOTOS Monte
  Carlo: A precision tool for QED corrections in $Z$ and $W$ decays}},
  }{}\href{http://dx.doi.org/10.1140/epjc/s2005-02396-4}{Eur.\ Phys.\ J.\
  \textbf{C45} (2006) 97}, \href{http://arxiv.org/abs/hep-ph/0506026}{{\tt
  arXiv:hep-ph/0506026}}\relax
\mciteBstWouldAddEndPuncttrue
\mciteSetBstMidEndSepPunct{\mcitedefaultmidpunct}
{\mcitedefaultendpunct}{\mcitedefaultseppunct}\relax
\EndOfBibitem
\bibitem{Allison:2006ve}
Geant4 collaboration, J.~Allison {\em et~al.},
  \ifthenelse{\boolean{articletitles}}{\emph{{Geant4 developments and
  applications}}, }{}\href{http://dx.doi.org/10.1109/TNS.2006.869826}{IEEE
  Trans.\ Nucl.\ Sci.\  \textbf{53} (2006) 270}\relax
\mciteBstWouldAddEndPuncttrue
\mciteSetBstMidEndSepPunct{\mcitedefaultmidpunct}
{\mcitedefaultendpunct}{\mcitedefaultseppunct}\relax
\EndOfBibitem
\bibitem{Agostinelli:2002hh}
Geant4 collaboration, S.~Agostinelli {\em et~al.},
  \ifthenelse{\boolean{articletitles}}{\emph{{Geant4: a simulation toolkit}},
  }{}\href{http://dx.doi.org/10.1016/S0168-9002(03)01368-8}{Nucl.\ Instrum.\
  Meth.\  \textbf{A506} (2003) 250}\relax
\mciteBstWouldAddEndPuncttrue
\mciteSetBstMidEndSepPunct{\mcitedefaultmidpunct}
{\mcitedefaultendpunct}{\mcitedefaultseppunct}\relax
\EndOfBibitem
\bibitem{LHCb-PROC-2011-006}
M.~Clemencic {\em et~al.}, \ifthenelse{\boolean{articletitles}}{\emph{{The
  \lhcb simulation application, Gauss: design, evolution and experience}},
  }{}\href{http://dx.doi.org/10.1088/1742-6596/331/3/032023}{{J.\ Phys.\ Conf.\
  Ser.\ } \textbf{331} (2011) 032023}\relax
\mciteBstWouldAddEndPuncttrue
\mciteSetBstMidEndSepPunct{\mcitedefaultmidpunct}
{\mcitedefaultendpunct}{\mcitedefaultseppunct}\relax
\EndOfBibitem
\bibitem{PDG2014}
Particle Data Group, K.~A. Olive {\em et~al.},
  \ifthenelse{\boolean{articletitles}}{\emph{{\href{http://pdg.lbl.gov/}{Review
  of particle physics}}},
  }{}\href{http://dx.doi.org/10.1088/1674-1137/38/9/090001}{Chin.\ Phys.\
  \textbf{C38} (2014) 090001}\relax
\mciteBstWouldAddEndPuncttrue
\mciteSetBstMidEndSepPunct{\mcitedefaultmidpunct}
{\mcitedefaultendpunct}{\mcitedefaultseppunct}\relax
\EndOfBibitem
\bibitem{Pivk:2004ty}
M.~Pivk and F.~R. Le~Diberder,
  \ifthenelse{\boolean{articletitles}}{\emph{{sPlot: a statistical tool to
  unfold data distributions}},
  }{}\href{http://dx.doi.org/10.1016/j.nima.2005.08.106}{Nucl.\ Instrum.\
  Meth.\  \textbf{A555} (2005) 356},
  \href{http://arxiv.org/abs/physics/0402083}{{\tt
  arXiv:physics/0402083}}\relax
\mciteBstWouldAddEndPuncttrue
\mciteSetBstMidEndSepPunct{\mcitedefaultmidpunct}
{\mcitedefaultendpunct}{\mcitedefaultseppunct}\relax
\EndOfBibitem
\bibitem{LHCb-PAPER-2013-003}
LHCb collaboration, R.~Aaij {\em et~al.},
  \ifthenelse{\boolean{articletitles}}{\emph{{Search for direct $CP$ violation
  in $D^0 \to h^- h^+$ modes using semileptonic $B$ decays}},
  }{}\href{http://dx.doi.org/10.1016/j.physletb.2013.04.061}{Phys.\ Lett.\
  \textbf{B723} (2013) 33}, \href{http://arxiv.org/abs/1303.2614}{{\tt
  arXiv:1303.2614}}\relax
\mciteBstWouldAddEndPuncttrue
\mciteSetBstMidEndSepPunct{\mcitedefaultmidpunct}
{\mcitedefaultendpunct}{\mcitedefaultseppunct}\relax
\EndOfBibitem
\bibitem{LHCb-PAPER-2014-053}
LHCb collaboration, R.~Aaij {\em et~al.},
  \ifthenelse{\boolean{articletitles}}{\emph{{Measurement of the semileptonic
  \CP asymmetry in \Bz--\Bzb mixing}},
  }{}\href{http://dx.doi.org/10.1103/PhysRevLett.114.041601}{Phys.\ Rev.\
  Lett.\  \textbf{114} (2015) 041601},
  \href{http://arxiv.org/abs/1409.8586}{{\tt arXiv:1409.8586}}\relax
\mciteBstWouldAddEndPuncttrue
\mciteSetBstMidEndSepPunct{\mcitedefaultmidpunct}
{\mcitedefaultendpunct}{\mcitedefaultseppunct}\relax
\EndOfBibitem
\bibitem{LHCb-PAPER-2014-042}
LHCb collaboration, R.~Aaij {\em et~al.},
  \ifthenelse{\boolean{articletitles}}{\emph{{Measurement of the
  $\bar{B}^0-B^0$ and $\bar{B}^0_s-B^0_s$ production asymmetries in $pp$
  collisions at $\sqrt{s}=7$ TeV}},
  }{}\href{http://dx.doi.org/10.1016/j.physletb.2014.10.005}{Phys.\ Lett.\
  \textbf{B739} (2014) 218}, \href{http://arxiv.org/abs/1408.0275}{{\tt
  arXiv:1408.0275}}\relax
\mciteBstWouldAddEndPuncttrue
\mciteSetBstMidEndSepPunct{\mcitedefaultmidpunct}
{\mcitedefaultendpunct}{\mcitedefaultseppunct}\relax
\EndOfBibitem
\bibitem{LHCb-DP-2014-001}
R.~Aaij {\em et~al.}, \ifthenelse{\boolean{articletitles}}{\emph{{Performance
  of the LHCb Vertex Locator}},
  }{}\href{http://dx.doi.org/10.1088/1748-0221/9/09/P09007}{JINST \textbf{9}
  (2014) P09007}, \href{http://arxiv.org/abs/1405.7808}{{\tt
  arXiv:1405.7808}}\relax
\mciteBstWouldAddEndPuncttrue
\mciteSetBstMidEndSepPunct{\mcitedefaultmidpunct}
{\mcitedefaultendpunct}{\mcitedefaultseppunct}\relax
\EndOfBibitem
\bibitem{LHCb-PAPER-2013-006}
LHCb collaboration, R.~Aaij {\em et~al.},
  \ifthenelse{\boolean{articletitles}}{\emph{{Precision measurement of the
  \Bs--\Bsb oscillation frequency in the decay $B^0_s \to D^-_s \pi^+$}},
  }{}\href{http://dx.doi.org/10.1088/1367-2630/15/5/053021}{New J.\ Phys.\
  \textbf{15} (2013) 053021}, \href{http://arxiv.org/abs/1304.4741}{{\tt
  arXiv:1304.4741}}\relax
\mciteBstWouldAddEndPuncttrue
\mciteSetBstMidEndSepPunct{\mcitedefaultmidpunct}
{\mcitedefaultendpunct}{\mcitedefaultseppunct}\relax
\EndOfBibitem
\bibitem{Neyman}
J.~Neyman, \ifthenelse{\boolean{articletitles}}{\emph{{Outline of a theory of
  statistical estimation based on the classical theory of probability}},
  }{}\href{http://dx.doi.org/10.1098/rsta.1937.0005}{Phil.\ Trans.\ R.\ Soc.\
  \textbf{A236} (1937) 333}\relax
\mciteBstWouldAddEndPuncttrue
\mciteSetBstMidEndSepPunct{\mcitedefaultmidpunct}
{\mcitedefaultendpunct}{\mcitedefaultseppunct}\relax
\EndOfBibitem
\end{mcitethebibliography}

\newpage

% Author List ----------------------------  

%%%%%%%%%%%%%%%%%%%%%%%%%%%%%%%%%%%%%%%%%%
\centerline{\large\bf LHCb collaboration}
\begin{flushleft}
\small
R.~Aaij$^{41}$, 
B.~Adeva$^{37}$, 
M.~Adinolfi$^{46}$, 
A.~Affolder$^{52}$, 
Z.~Ajaltouni$^{5}$, 
S.~Akar$^{6}$, 
J.~Albrecht$^{9}$, 
F.~Alessio$^{38}$, 
M.~Alexander$^{51}$, 
S.~Ali$^{41}$
% ...

{\footnotesize \it
% ...
$ ^{5}$Clermont Universit\'{e}, Universit\'{e} Blaise Pascal, CNRS/IN2P3, LPC, Clermont-Ferrand, France\\
$ ^{6}$CPPM, Aix-Marseille Universit\'{e}, CNRS/IN2P3, Marseille, France\\
% ...
$ ^{9}$Fakult\"{a}t Physik, Technische Universit\"{a}t Dortmund, Dortmund, Germany\\
% ...
$ ^{37}$Universidad de Santiago de Compostela, Santiago de Compostela, Spain\\
$ ^{38}$European Organization for Nuclear Research (CERN), Geneva, Switzerland\\
% ...
$ ^{41}$Nikhef National Institute for Subatomic Physics, Amsterdam, The Netherlands\\
% ...
$ ^{46}$H.H. Wills Physics Laboratory, University of Bristol, Bristol, United Kingdom\\
% ...
$ ^{51}$School of Physics and Astronomy, University of Glasgow, Glasgow, United Kingdom\\
$ ^{52}$Oliver Lodge Laboratory, University of Liverpool, Liverpool, United Kingdom\\
% ...
}
\end{flushleft}
%%%%%%%%%%%%%%%%%%%%%%%%%%%%%%%%%%%%%%%%%%

\end{document}